\documentclass[journal,twoside,web]{ieeecolor}
\usepackage{jsen}
\usepackage{cite}
\usepackage{amsmath,amssymb,amsfonts}
\usepackage{graphicx}
\usepackage{textcomp}
\usepackage{wrapfig}

\usepackage{amsmath,amssymb,amsfonts}
\usepackage{algorithm}
\usepackage{algorithmicx}
\usepackage[noend]{algpseudocode} % no end hides endif endfor
\usepackage{subcaption}
\usepackage{xcolor}
\usepackage{commath} % to use abs

\usepackage{booktabs}   % For professional horizontal lines (\toprule, \midrule, \bottomrule)
\usepackage{tabularx}  % For adjustable width columns
\usepackage{pifont}    % For checkmarks and crosses
\usepackage{xcolor}    % For colored text if needed

\newcommand{\cmark}{\ding{51}} % Checkmark
\newcommand{\xmark}{\ding{55}} % X mark

\def\BibTeX{{\rm B\kern-.05em{\sc i\kern-.025em b}\kern-.08em
    T\kern-.1667em\lower.7ex\hbox{E}\kern-.125emX}}
\definecolor{abstractbg}{rgb}{0.89804,0.94510,0.83137}
\usepackage{tikz}

\begin{document}
%\the\textwidth 516 pt
%\the\linewidth 252 pt
\title{UAV-Based 3D Spectrum Sensing: Insights on Altitude, Bandwidth, Trajectory, and Effective Antenna Patterns on REM Reconstruction}
\author{Mushfiqur Rahman, \IEEEmembership{Graduate Student Member, IEEE}, Sung Joon Maeng, \IEEEmembership{Member, IEEE}, \.{I}smail G\"{u}ven\c{c}, \IEEEmembership{Fellow, IEEE}, Chau-Wai Wong, \IEEEmembership{Senior Member, IEEE}, Mihail Sichitiu, \IEEEmembership{Senior Member, IEEE}, Jason A. Abrahamson, and Arupjyoti Bhuyan, \IEEEmembership{Senior Member, IEEE}
%\thanks{Received. \textit{(Corresponding author: {I}smail~G\"{u}ven\c{c}.)}}
\thanks{This paper has been submitted for possible publication to IEEE. Copyright may be transferred without notice.}
\thanks{Please see the Acknowledgment section of this article for the funding and author affiliations. \textit{(Corresponding author: {I}smail~G\"{u}ven\c{c}.)}}
\thanks{The code, dataset, and reproducible results are available at https://github.com/MPACT-Lab/uav-spectrum-sensing-insights/.}
}

\IEEEtitleabstractindextext{%
\fcolorbox{abstractbg}{abstractbg}{%
\begin{minipage}{\textwidth}%
\begin{wrapfigure}[16]{r}{2.8in}%
\includegraphics[width=2.6in]{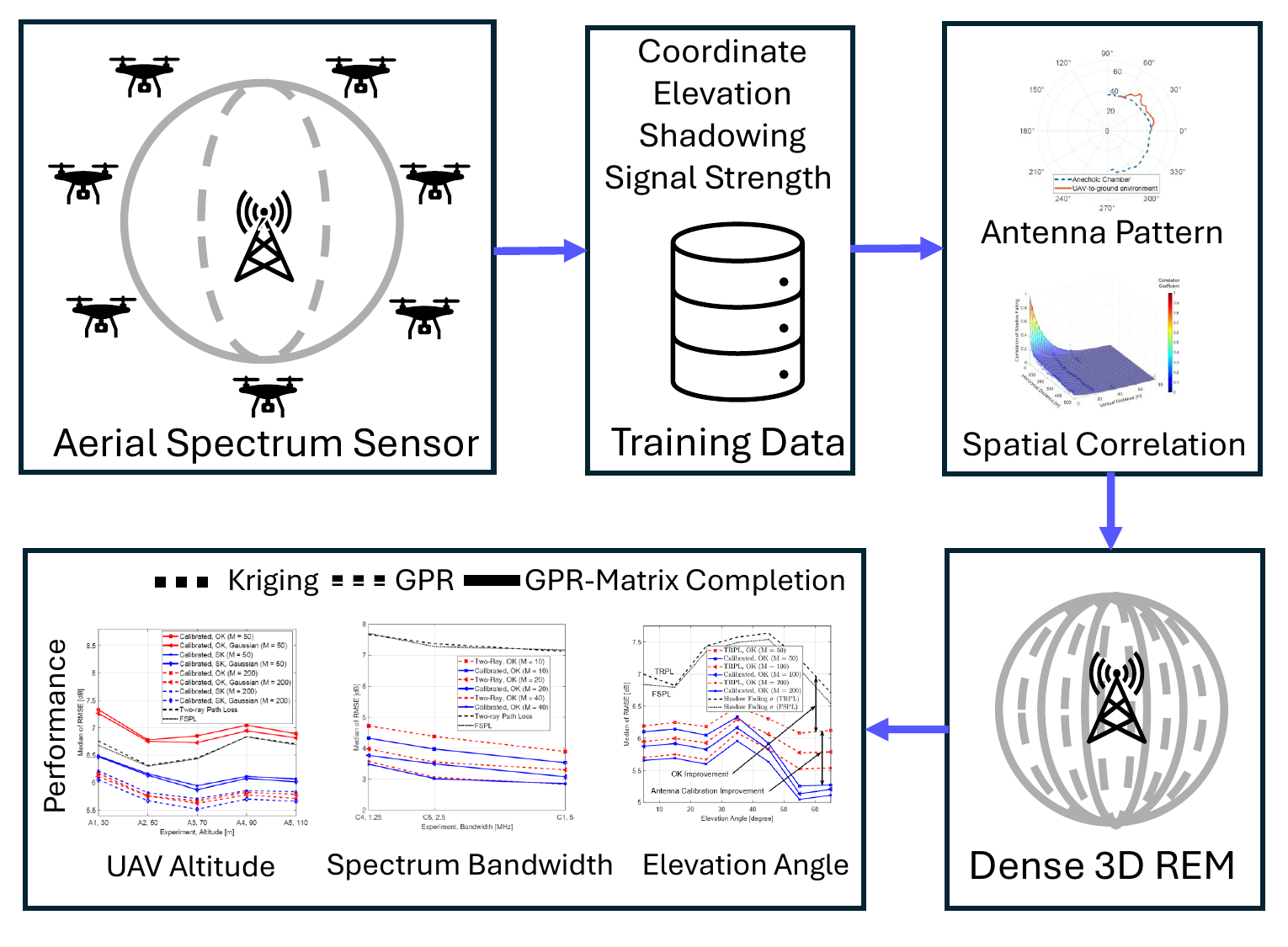}%
\end{wrapfigure}%
\begin{abstract}%
Spectrum sensing and the generation of 3D Radio Environment Maps (REMs) are essential for enabling spectrum sharing within cognitive radio networks. While Uncrewed Aerial Vehicles (UAVs) offer high-mobility 3D sensing, REM accuracy is challenged by dynamic flight behaviors, where fluctuations in UAV speed and direction introduce measurement inconsistencies. Furthermore, the airframe itself impacts the onboard antenna's radiation characteristics.
In this paper, using real-world data, we systematically analyze how REM reconstruction accuracy is shaped by three key pillars: physical sensing parameters like altitude and bandwidth, environmental shadowing, and distortions caused by the UAV airframe.
First, we benchmark diverse spatial prediction models, including simple Kriging (SK), ordinary Kriging (OK), trans-Gaussian Kriging, and Gaussian process regression (GPR). We demonstrate that while SK and its trans-Gaussian variant are highly accurate at extreme sample sparsity, OK improves as sample size increases, and GPR serves as the most stable overall baseline.
Building on this, we propose a novel matrix completion (MC)-assisted GPR framework that enhances REM reconstruction in the presence of non-uniform spatial smoothness. The method operates by decomposing the REM into two distinct layers: a global smooth component and a highly varying local component.
Our analysis based on real-world measurements reveals three key findings: 1) REM accuracy and shadowing variance follow a distinct tri-phasic trend as the UAV altitude increases; 2) REM accuracy significantly improves with increased spectrum bandwidth; and 3) antenna pattern calibration from in-field measurements significantly enhances REM accuracy by accounting for the effect of the UAV airframe.
\end{abstract}

\end{minipage}}}

\maketitle
%\copyrightnotice

\section{Introduction}
\IEEEPARstart{T}{he} rapid expansion of wireless ecosystems, spanning terrestrial and aerial domains, has necessitated a fundamental shift in how spectral resources are perceived and managed. As the density of diverse transmitters, including femtocells, Internet of Things (IoT) devices, 5G-enabled vehicles, and Uncrewed Aerial Vehicles (UAVs), continues to surge, the 3D spectrum environment has become increasingly congested~\cite{wang2023sparse}. Consequently, this complexity precipitates a critical state of spectrum scarcity where traditional static allocation models fall short of meeting the demands of modern connectivity. To address these challenges, cognitive radio technology leverages intelligent spectrum management strategies, such as dynamic spectrum sharing, which relies on comprehensive, real-time awareness of spectrum occupancy~\cite{zhou2023accurate}. Central to this awareness is the 3D Radio Environment Map (REM), a spatial database that characterizes received power levels, such as Reference Signal Received Power (RSRP), across specific coordinates and altitudes. Constructing a dense, high-fidelity 3D REM is therefore a foundational requirement for optimizing power control and frequency allocation while ensuring interference mitigation in a multi-layered wireless landscape.

Constructing a comprehensive 3D REM begins with the acquisition of RSRP measurements across diverse spatial locations. As exhaustive sampling is physically prohibitive, the process of REM generation relies on predictive strategies, such as spatial interpolation and channel modeling, to estimate signal levels in unobserved regions~\cite{du2021uav}. Recent advancements in REM reconstruction~\cite{shen2019uav, wei2022three, romero2020aerial, shrestha2022spectrum, du2021uav, ivanov2022uav, zhu2022demo, zeng2022uav} have increasingly leveraged UAVs to facilitate these measurements, utilizing both fixed and autonomous trajectories in 3D environments. GPS-integrated, lightweight UAVs provide the agility required to navigate challenging terrains and precisely reach high-altitude coordinates that are inaccessible to ground-based infrastructure.
Furthermore, the literature has explored sophisticated UAV trajectory planning algorithms designed to enhance REM fidelity while minimizing operational overhead, such as flight time~\cite{romero2020aerial, shrestha2022spectrum, wei2022three}. These approaches often employ Bayesian uncertainty metrics or deep learning frameworks to direct UAVs toward the most informative sampling locations, effectively prioritizing areas with high spatial variance or significant shadow fading (SF).

Despite the distinct advantages of UAV-assisted sensing, several platform-specific challenges complicate the acquisition of reliable data. First, real-world UAV measurements are inherently susceptible to diverse noise sources, including electromagnetic interference from rotors and onboard electronics, which are compounded by environmental SF and multi-path fading; consequently, reconstruction methods validated solely on statistical or ray-tracing-based synthetic data may not accurately reflect performance in practical deployments. Second, the spatial correlation of measurements is heavily influenced by the flight path; for instance, traversing the same trajectory from opposite directions or at varying speeds can yield significant signal variations due to changes in UAV orientation and heading. These kinematic factors, combined with the typically limited nature of real-world datasets, necessitate rigorous validation protocols tailored specifically for UAV-based sensing. Finally, the physical airframe of the UAV can significantly alter the antenna's radiation characteristics compared to its baseline performance in an anechoic chamber, introducing gain distortions.

Efficient sample acquisition and intelligent REM reconstruction are the two vital components in achieving a high-fidelity REM. Existing literature addresses sample acquisition through data-intensive UAV trajectory optimization algorithms. However, critical gaps remain in characterizing the impact of physical parameters, such as altitude, elevation angle, and spectrum bandwidth, on the information richness of a measurement sample. Current REM reconstruction methodologies suffer from two critical gaps: 1) they assume uniform spatial smoothness across the entire map, and 2) they do not account for the impact of the UAV airframe. To address these gaps, using real-world data, our contributions are as follows:
\begin{itemize}
    \item \textbf{Impact of Flight Altitude and Elevation Angle:} We analyze the SF variance and REM accuracy across varying flight altitudes and increasing elevation angles. This analysis uncovers distinct tri-phasic trends that directly link REM reconstruction accuracy to altitude and elevation angle. Identifying high-fluctuation zones enables the optimization of the amount of data collection across diverse altitudes and elevations.
    
    \item \textbf{Impact of Spectrum Sensing Bandwidth:}
    We evaluate REM reconstruction accuracy across a wide range of bandwidths. Our results demonstrate that broader sensing bandwidth consistently enhances REM accuracy by leveraging frequency diversity. These findings provide guidelines on the number of measurement samples needed for constructing an REM for a given bandwidth.
    
    \item \textbf{Two-Layer SF Map and Local Deep-Shadow Extraction:} We propose \textit{matrix completion~(MC)-assisted Gaussian process regression~(GPR)}, a two-layer reconstruction framework. Traditional methods, including ordinary Kriging~(OK), simple Kriging~(SK), and GPR, apply a fixed spatial correlation model throughout the map, equally smoothing all areas of a REM. In contrast, MC-assisted GPR isolates the zones of high spatial variance or local deep-shadow into a separate layer. This approach outperforms the baselines in REM accuracy.
    
    \item \textbf{In-Flight Antenna Pattern Calibration (APC):} We model the UAV airframe-induced electromagnetic interference and signal blockage as an effective antenna pattern, learning directly from in-flight empirical data. We integrate this calibrated pattern into the core deterministic propagation model, showing that the proposed approach significantly improves the REM reconstruction accuracy.
\end{itemize}

The rest of the paper is organized as follows. Section~\ref{sec:rel_works} outlines the literature review. Section~\ref{sec:sys_model} formally defines the REM reconstruction problem and introduces foundational two-ray path loss~(TRPL) and SF models. Section~\ref{sec:datasets} describes the datasets. Section~\ref{sec:sensing_method} reviews the baseline REM reconstruction algorithms and introduces the proposed MC-assisted GPR. Section~\ref{sec:calib_rad_prop} presents the APC methodology and empirical antenna pattern examples. Section~\ref{sec:num_results} demonstrates the experimental results, including the impact of altitude, elevation angle, and bandwidth on REM accuracy, and the final section concludes the paper.
\begin{table*}[!t]
\centering
\caption{Literature review of radio environment map reconstruction.}
\label{tab:lit-review}
\small % Slightly smaller font to fit all columns comfortably
\begin{tabularx}{\textwidth}{l X c c c c c c}
\toprule
\begin{tabular}[c]{@{}c@{}}Literature\\ (First author~\textit{et~al.}, year)\end{tabular} & Method & \begin{tabular}[c]{@{}c@{}c@{}}Input\\ (Tx Location,\\ Observations)\end{tabular} & \begin{tabular}[c]{@{}c@{}}Multi-\\ altitude\end{tabular} & \begin{tabular}[c]{@{}c@{}}Multi-\\ bandwidth\end{tabular} & \begin{tabular}[c]{@{}c@{}}UAV-based\\ measurements\end{tabular} & \begin{tabular}[c]{@{}c@{}}Elevation\\ dependency\end{tabular} & \begin{tabular}[c]{@{}c@{}}Radiation\\ calibration\end{tabular} \\ 
\midrule
Pesko (2015)~\cite{pesko2015indirect} & Okumura-Hata & \textcolor{green}{\cmark} \, \textcolor{green}{\cmark} &  \textcolor{red}{\xmark} &  \textcolor{red}{\xmark} &  \textcolor{green}{\cmark} &  \textcolor{red}{\xmark} &  \textcolor{green}{\cmark} \\
Krijestorac~(2021)~\cite{krijestorac2021spatial}& U-Net &  \textcolor{red}{\xmark} \,  \textcolor{green}{\cmark} &  \textcolor{red}{\xmark} &  \textcolor{red}{\xmark} &  \textcolor{red}{\xmark} &  \textcolor{red}{\xmark} &  \textcolor{red}{\xmark} \\ 
Levie~(2021)~\cite{levie2021radiounet}& U-Net (RadioUnet) &  \textcolor{green}{\cmark} \,  \textcolor{green}{\cmark} &  \textcolor{red}{\xmark} &  \textcolor{red}{\xmark} &  \textcolor{red}{\xmark} &  \textcolor{red}{\xmark} &  \textcolor{red}{\xmark} \\ 
Teganya~(2021)~\cite{teganya2021deep}& Autoencoder &  \textcolor{red}{\xmark} \,  \textcolor{green}{\cmark} &  \textcolor{red}{\xmark} &  \textcolor{red}{\xmark} &  \textcolor{red}{\xmark} &  \textcolor{red}{\xmark} &  \textcolor{red}{\xmark} \\ 
Mao~(2022)~\cite{mao2022machine} & BPNN, GAN &  \textcolor{green}{\cmark} \,  \textcolor{green}{\cmark} &  \textcolor{red}{\xmark} &  \textcolor{red}{\xmark} &  \textcolor{green}{\cmark} &  \textcolor{red}{\xmark} &  \textcolor{red}{\xmark} \\
Sun~(2022)~\cite{sun2022propagation} & MC &  \textcolor{red}{\xmark} \,  \textcolor{green}{\cmark} &  \textcolor{red}{\xmark} &  \textcolor{red}{\xmark} &  \textcolor{red}{\xmark} &  \textcolor{red}{\xmark} &  \textcolor{red}{\xmark}\\
Liu~(2023)~\cite{liu2023uav} & RT, Kriging &  \textcolor{green}{\cmark} \,  \textcolor{green}{\cmark} &  \textcolor{red}{\xmark} &  \textcolor{red}{\xmark} &  \textcolor{red}{\xmark} &  \textcolor{red}{\xmark} &  \textcolor{red}{\xmark} \\ 
Maeng~(2023)~\cite{maeng2023kriging} & TRPL, Kriging &  \textcolor{green}{\cmark} \,  \textcolor{green}{\cmark} &  \textcolor{red}{\xmark} &  \textcolor{red}{\xmark} &  \textcolor{green}{\cmark} &  \textcolor{red}{\xmark} &  \textcolor{red}{\xmark} \\
Hu~(2023)~\cite{hu20233d}& GAN &  \textcolor{red}{\xmark} \,  \textcolor{green}{\cmark} &  \textcolor{green}{\cmark} &  \textcolor{red}{\xmark} &  \textcolor{red}{\xmark} &  \textcolor{red}{\xmark} &  \textcolor{red}{\xmark} \\
Ivanov (2023)~\cite{ivanov2023limited} & IDW, Kriging &  \textcolor{red}{\xmark} \,  \textcolor{green}{\cmark} &  \textcolor{green}{\cmark} &  \textcolor{red}{\xmark} &  \textcolor{green}{\cmark} &  \textcolor{red}{\xmark} &  \textcolor{red}{\xmark} \\
Locke~(2023)~\cite{locke2023radio}& Autoencoder &  \textcolor{red}{\xmark} \,  \textcolor{green}{\cmark} &  \textcolor{red}{\xmark} &  \textcolor{red}{\xmark} &  \textcolor{red}{\xmark} &  \textcolor{red}{\xmark} &  \textcolor{red}{\xmark} \\ 
Zhang~(2023)~\cite{zhang2023rme}& GAN (RME-GAN) &  \textcolor{green}{\cmark} \,  \textcolor{green}{\cmark} &  \textcolor{red}{\xmark} &  \textcolor{red}{\xmark} &  \textcolor{red}{\xmark} &  \textcolor{red}{\xmark} &  \textcolor{red}{\xmark} \\ 
Chaves~(2023)~\cite{chaves2023deeprem}& U-Net, cGAN &  \textcolor{red}{\xmark} \,  \textcolor{green}{\cmark} &  \textcolor{red}{\xmark} &  \textcolor{red}{\xmark} &  \textcolor{red}{\xmark} &  \textcolor{red}{\xmark} &  \textcolor{red}{\xmark} \\
Zeng~(2023)~\cite{zeng2022uav}& CNN &  \textcolor{green}{\cmark} \,  \textcolor{green}{\cmark} &  \textcolor{red}{\xmark} &  \textcolor{red}{\xmark} &  \textcolor{red}{\xmark} &  \textcolor{red}{\xmark} &  \textcolor{red}{\xmark} \\
Wang~(2024)~\cite{wang2024sparse} & SBL, GPR &  \textcolor{green}{\cmark} \,  \textcolor{green}{\cmark} &  \textcolor{green}{\cmark} &  \textcolor{red}{\xmark} &  \textcolor{red}{\xmark} &  \textcolor{red}{\xmark} &  \textcolor{red}{\xmark} \\
Rahman~(2024)~\cite{rahman20243d}& TRPL, Kriging, MC &  \textcolor{green}{\cmark} \,  \textcolor{green}{\cmark} &  \textcolor{red}{\xmark} &  \textcolor{red}{\xmark} &  \textcolor{green}{\cmark} &  \textcolor{red}{\xmark} &  \textcolor{red}{\xmark} \\ 
Wang~(2024)~\cite{wang2024radiodiff}& Diffusion (RadioDiff) &  \textcolor{green}{\cmark} \,  \textcolor{red}{\xmark} &  \textcolor{red}{\xmark} &  \textcolor{red}{\xmark} &  \textcolor{red}{\xmark} &  \textcolor{red}{\xmark} &  \textcolor{red}{\xmark} \\ 
Alobaidy~(2025)~\cite{alobaidy2025empirical} & SLR, EBT, GPR &  \textcolor{red}{\xmark} \,  \textcolor{green}{\cmark} &  \textcolor{green}{\cmark} &  \textcolor{red}{\xmark} &  \textcolor{green}{\cmark} &  \textcolor{red}{\xmark} &  \textcolor{red}{\xmark} \\ 
\midrule
\textbf{This paper} & TRPL, GPR, MC & \textbf{ \textcolor{green}{\cmark}} \textbf{ \textcolor{green}{\cmark}} &  \textcolor{green}{\cmark} &  \textcolor{green}{\cmark} &  \textcolor{green}{\cmark} &  \textcolor{green}{\cmark} &  \textcolor{green}{\cmark} \\
\bottomrule
\end{tabularx}
\end{table*}

\section{Related Work}
\label{sec:rel_works}
This section first reviews the evolution of reconstruction methodologies from traditional data-driven interpolation to modern hybrid frameworks. We then examine the existing literature on altitude-dependent REM reconstruction and UAV airframe-induced antenna distortions.
\subsection{REM Reconstruction Methodologies}
The reconstruction of REMs from sparse UAV-collected measurements remains a significant research challenge. Existing literature generally categorizes these methodologies into three primary frameworks: direct, indirect, and hybrid reconstruction techniques~\cite{gajewski2018propagation}.

\textbf{Direct REM Reconstruction Methods} treat the problem as a generic data interpolation or tensor completion task, without incorporating the underlying physics of radio propagation. This category includes classical spatial interpolation methods such as Kriging~\cite{sato2017Kriging}, Inverse Distance Weighting (IDW), and GPR. GPR is typically implemented in two forms: 1) \textit{spatial GPR}, which is mathematically analogous to Kriging~\cite{wang2024sparse}, and 2) \textit{feature-based GPR}, which utilizes variables such as 3D distance, azimuth, and elevation to predict received power~\cite{alobaidy2025empirical}. In this paper, we employ the first form of GPR, which extends simple Kriging by explicitly modeling measurement noise as a Gaussian process. The efficacy of Kriging and spatial GPR depends on the covariance kernel, which models spatial signal correlation. Common kernel selections include a family of exponential, Matern, and sinusoidal exponential functions. While kernel hyperparameters are traditionally estimated from sampled data, calibrating them in a controlled anechoic chamber has also been proposed~\cite{ivanov2024calibration}, as data anomalies can hamper their performance~\cite{farnham2016radio, xia2020radio}.

Deep Learning (DL) models have gained traction due to their superior generalization capabilities, despite the requirement for extensive training datasets~\cite{chaves2023deeprem, zeng2022uav, ivanov2024deep, reddy2026transforem}. Current DL architectures for REM generation primarily utilize U-Net~\cite{levie2021radiounet, krijestorac2021spatial}, autoencoders~\cite{locke2023radio, teganya2021deep}, Generative Adversarial Networks (GANs)~\cite{zhang2023rme, hu20233d}, or diffusion models~\cite{wang2024radiodiff}. These methodologies are further divided into \textit{sampling-free} approaches, which rely solely on transmitter (Tx) locations and environmental metadata (e.g., terrain or building maps) as their input, and \textit{sampling-based} approaches that integrate sparse measurements. This work specifically addresses scenarios involving known Tx locations and limited sampled measurements.
Notably, most of the state-of-the-art DL methods are trained and validated on synthetic data rather than in-field UAV measurements. A recent study identified the U-Net architecture as the top-performing model, utilizing the extensive \textit{SpectrumNet} simulated radio map dataset~\cite{zhang2024generative}. Finally, MC~\cite{sun2022propagation, zhang2020spectrum} leverages low-rank characteristics to predict unknown grid values. While reliable for single-source scenarios, its performance often degrades due to oversmoothing in multi-source or deep-shadowed environments~\cite{wang2025robust}.

\textbf{Indirect REM Reconstruction Methods} leverage the physics of radio propagation by assuming a deterministic propagation model, such as log-distance or TRPL, and estimating model parameters from observed samples. These methods are termed ``indirect" because the REM is generated via a tuned channel model. The process involves estimating variables such as the path loss exponent, line-of-sight~(LoS) probability, SF variance, source location, and antenna radiation patterns~\cite{gajewski2018propagation, masrur2025bridging}. While pure modeling (e.g., ray-tracing) can be performed without measurements, indirect methods utilize sparse samples to adapt the model to specific site characteristics. For instance, Pesko~\textit{et al.}~\cite{pesko2015indirect} utilized the Okumura-Hata model to estimate source location and tune Tx antenna patterns via non-linear optimization. Other studies address source localization and parameter estimation using Orthogonal Matching Pursuit (OMP)~\cite{shen20213d} or Sparse Bayesian Learning (SBL)~\cite{wang2023sparse}. While these approaches eliminate the need to store large volumes of sparse samples, their accuracy is inherently limited by how well the chosen channel model reflects the underlying RF distribution of the real-world environment.

\textbf{Hybrid Methods} bridge the gap between direct and indirect frameworks. While direct methods are model-agnostic, indirect methods provide physical grounding; a hybrid approach typically yields superior accuracy by combining these strengths~\cite{tsukamoto2018highly, du2021uav}. For example, Mao~\textit{et~al.}~\cite{mao2022machine} employed a geometric propagation model to identify multipath delays, subsequently using a machine learning model to predict individual path powers. Similarly, Liu~\textit{et~al.}~\cite{liu2023uav} modeled deterministic path loss using environmental semantics and estimated residual SF via Kriging. Likewise, Maeng~\textit{et~al.}~\cite{maeng2023kriging} employed TRPL and Kriging. In our work, we adopt TRPL with GPR. Given that the transmitter location is known, we first utilize TRPL to establish the large-scale path loss trend. Subsequently, we model the residual SF through GPR to capture localized environmental characteristics that are otherwise difficult to characterize through purely physics-based models.

\subsection{Altitude-Dependent REM Reconstruction}
Ivanov~\textit{et~al.}~\cite{ivanov2023limited} evaluated Kriging and IDW performance across four distinct UAV altitudes. Although the authors did not explicitly analyze the altitude-dependency of their results, a secondary analysis of their data reveals a distinct tri-phasic trend: for both methodologies, the REM reconstruction error initially decreases as the altitude rises from 80~m to 90~m, undergoes a localized increase at 100~m, and subsequently decreases at 110~m. Maeng~\textit{et~al.}~\cite{maeng2023kriging} analyzed REM reconstruction error using multi-altitude UAV data by varying the vertical separation between training and testing sets (i.e., altitude offsets of 0~m, 10~m, and 20~m). Our objective diverges from~\cite{maeng2023kriging} in that we investigate the intrinsic difficulty of intra-altitude prediction at each specific height layer, whereas the focus in~\cite{maeng2023kriging} was to quantify the degradation in accuracy as the vertical distance between training and testing data increases.

\subsection{Airframe-Induced Radiation Pattern Distortions}
Recent studies~\cite{badi2019experimental,badi2020experimentally, semkin2021lightweight} experimentally demonstrated that the antenna radiation pattern is significantly altered when attached to a UAV, primarily due to uneven reflections and scattering, which introduce additional multipath components. Similarly, Sun~\textit{et~al.}~\cite{sun2017air} showed that the UAV wings can cause airframe shadowing, particularly at low elevation angles, such as when the UAV is slightly tilted, blocking the direct ray. Despite these known physical impairments, the impact of UAV-specific APC on the accuracy of 3D REM reconstruction remains underexplored. While Pesko~\textit{et~al.}~\cite{pesko2015indirect} demonstrated that APC can effectively reduce reconstruction errors, their findings were based on simulated data. Furthermore, Maeng~\textit{et~al.}~\cite{maeng2023kriging} utilized anechoic chamber measurements to characterize the antenna in isolation; however, this approach does not account for the pattern distortion caused by the UAV airframe itself. While it is possible to measure the composite UAV-antenna pattern in a large-scale anechoic facility~\cite{semkin2021lightweight}, in this work, we explore the feasibility of estimating the effective antenna pattern directly from in-field empirical measurements. This allows the model to better capture the interaction between the antenna and the UAV airframe while in operation.

To the best of our knowledge, the dependence of REM reconstruction error on altitude, bandwidth, and elevation angle, alongside the integration of APC using empirical UAV-collected data, remains insufficiently addressed in existing literature. Table~\ref{tab:lit-review} provides a comparative summary, highlighting the distinct contributions of this study relative to the current state-of-the-art in REM reconstruction.
\begin{table}[htbp]
\centering
\caption{Key abbreviations.}
\label{tab:abbr}
\small % Slightly smaller font to fit all columns comfortably
\begin{tabularx}{\linewidth}{l l}
\toprule
Abbreviation & Full name \\ 
\midrule
APC & Antenna Pattern Calibration\\
BPNN & Back-Propagation Neural Network\\
BS & Base Station\\
GS & Ground Station\\
CNN & Convolutional Neural Network\\
EBT & Ensemble of Bagged Trees\\
FSPL  & Free-Space Path Loss \\
GAN  & Generative Adversarial Network \\
cGAN  & Conditional Generative Adversarial Network \\
GPR  & Gaussian Process Regression \\
IDW  & Inverse Distance Weighting \\
LoS & Line of Sight\\
TRPL  & Two-Ray Path Loss \\
MC  & Matrix Completion \\
RT & Ray Tracing\\
REM & Radio Environment Maps\\
SBL & Sparse Bayesian Learning\\
%SD & Standard Deviation\\
SF & Shadow Fading\\
SLR & Stepwise Linear Regression\\
Tx & Transmitter\\
Rx & Receiver\\
UGV & Unmanned Ground Vehicle\\
\bottomrule
\end{tabularx}
\end{table}

\begin{table}[htbp]
\centering
\caption{Key variables in this study.}
\label{tab:var_list}
\small % Slightly smaller font to fit all columns comfortably
\begin{tabularx}{\linewidth}{l l}
\toprule
Symbol & Definition \\ 
\midrule
\(\theta\) & Elevation angle of the UAV\\
\(\theta_\text{t}\) & Elevation angle of the LoS path at the GS\\
\(\theta_\text{r}\) & Elevation angle of the LoS path at the UAV\\
\(\phi_\text{t}\) & Azimuth angle of the LoS path at the GS\\
\(\phi_\text{r}\) & Azimuth angle of the LoS path at the UAV\\
\(h_\text{uav}, H\) & Altitude of the UAV\\
\(h_\text{gs}\) & Altitude of the GS antenna\\
\(d_\text{h}\) & Horizontal distance between UAV and GS antenna\\
\(d_\text{v}\) & Vertical distance between GS and UAV antenna\\
\(d_\text{3D}\) & 3D distance between UAV and GS antenna\\
\(l_\text{gs}\) & 3D coordinate of the GS\\
\(l_\text{uav}\) & 3D coordinate of the UAV\\
\(\psi_{\mathrm{uav}}\) & Latitude of the UAV\\
\(\zeta_{\mathrm{uav}}\) & Longitude of the UAV\\
\(\psi_{\mathrm{gs}}\) & Latitude of the GS\\
\(\zeta_{\mathrm{gs}}\) & Longitude of the GS\\
\(\mathrm{P}_{\mathrm{Tx}}\) & Transmit power (known)\\
\(\mathrm{P}_{\mathrm{Rx}}\) & Received power (measured)\\
\(\hat{r}_{\mathrm{trpl}}\) & TRPL-estimated received power\\
%\(\hat{r}_{\mathrm{fspl}}\) & FSPL-estimated received power\\
%\(\hat{r}_{\mathrm{trpl}}^{(\text{APC})}\) & \begin{tabular}[l]{@{}l@{}}TRPL-estimated received power with antenna\\pattern calibration\end{tabular}\\
\(\hat{r}_{\mathrm{trpl}}^{(\text{APC})}\) & TRPL-estimated received power with APC\\
\(w_{\mathrm{trpl}},Z\) & \begin{tabular}[l]{@{}l@{}}SF component extracted as the RSRP residual\\ relative to TRPL\end{tabular}\\
\(\sigma_Z, \sigma_w, \sigma\) & Standard deviation of SF\\
\(R_w,R_Z\) & Spatial correlation of SF for a pair of locations\\
\(\mathbf{\gamma}\) & Semivariogram of SF for a pair of locations\\
\(C\) & Covariance of SF for a pair of locations\\
\(l_0\) & 3D coordinate of prediction location\\
\(\hat{w}, \hat{Z}\) & Estimated SF at prediction location\\
\(\mathrm{M}\) & Number of sparse samples available\\
\(\mathrm{R}\) & Sparse sample selection radius\\
\(\hat{r}\) & Estimated RSRP at prediction location\\
$\mathcal{E}$   & Test dataset (with $\mathrm{M}$ sparse samples) \\
$\mathcal{E}'$  & Auxiliary training dataset\\
\bottomrule
\end{tabularx}
\end{table}

\section{System Model Preliminaries}
\label{sec:sys_model}
This section establishes the basis of the analytical framework for 3D REM reconstruction. We first characterize the radio environment through a physics-based path loss model and a stochastic formulation of SF. Subsequently, we define the problem setup, which details the available system inputs, constraints, and the target reconstruction objectives.
\subsection{System Environment and Radio Propagation Model}
\begin{figure}% 
    \centering
		\vspace{-0mm}
    \includegraphics[width=\linewidth]{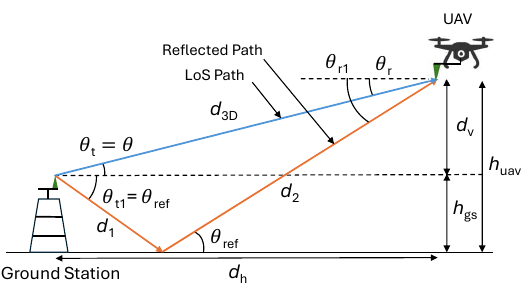}
    \hspace{4mm}
    \vspace{-3mm}
    \caption{Illustration of the Two-Ray Path Loss (TRPL) radio propagation model for an air-to-ground channel between horizontal rod-mounted antennas.}
	\vspace{-2mm}
    \label{fig:diagram_two-ray-model}
    \vspace{-1mm}
\end{figure}

The spectrum sensing environment in this study consists of two nodes: a stationary ground-based node and a portable node equipped with a UAV. The stationary node may be either a base station~(BS) or any other ground station~(GS), with the Radio Frequency (RF) front-end and communication modules positioned atop. The ground-based node emits the radio signal while the UAV collects RSRP measurements to sense the surrounding radio environment. The communication modules are mounted beneath the UAV, with the antenna oriented downward. Both nodes are equipped with antennas featuring nearly omnidirectional radiation patterns, as detailed in their datasheets. To minimize interference from the frame structure (e.g., the UAV body) and other RF communication modules, both antennas are affixed to horizontal rods, creating separation from other components of the UAV or GS. The environment is assumed to be rural, featuring a LoS and a ground-reflected ray, as depicted in Fig.~\ref{fig:diagram_two-ray-model}.

We define the 3D coordinates of the GS antenna $l_{\mathrm{gs}} = \left(\psi_{\mathrm{gs}}, \zeta_{\mathrm{gs}}, h_{\mathrm{gs}}\right)$ and the 3D coordinates of the UAV $l_{\mathrm{uav}} = \left(\psi_{\mathrm{uav}}, \zeta_{\mathrm{uav}}, h_{\mathrm{uav}}\right)$, where $\psi$, $\zeta$ and $h$ represent the latitude, longitude, and altitude, respectively. The horizontal and vertical distances between the GS and the UAV are:
\begin{subequations}
\begin{align}
d_{\mathrm{h}}\left(l_{\mathrm{gs}}, l_{\mathrm{uav}}\right) & = R_\text{e} \cdot \arccos \Big(\sin \psi_{\mathrm{uav }} \sin \psi_{\mathrm{gs}}\Big. \nonumber\\
&\Big.+\cos \psi_{\mathrm{uav }} \cos \psi_{\mathrm{gs}} \cos (\zeta_{\mathrm{gs}}-\zeta_{\mathrm{uav }})\Big) \label{eqn:dh},\\
d_{\mathrm{v}}\left(l_{\mathrm{gs}}, l_{\mathrm{uav}}\right) & =\big|h_{\mathrm{gs}}-h_{\mathrm{uav }}\big|,
\end{align}
\end{subequations}
where $R_\text{e}$ denotes the Earth's radius. The 3D distance, $d_{3\text{D}}(l_{\mathrm{gs}}, l_{\mathrm{uav}})$, is given by $\sqrt{d_\text{h}^2+d_\text{v}^2}$.
The TRPL model calculates the signal attenuation factor by considering both the LoS and ground-reflected paths as follows:
\begin{equation}
\begin{aligned}
\mathrm{A}_{\text {trpl}}&\left(l_{\mathrm{gs}}, l_{\mathrm{uav}}\right)=\left(\frac{l_{\text{wave}}}{4 \pi}\right)^2  \, \biggr \lvert \underbrace{\frac{\sqrt{\mathrm{G}_{\mathrm{gs}}\left(\phi_\text{t}, \theta_\text{t}\right) \mathrm{G}_{\mathrm{uav}}\left(\phi_\text{r}, \theta_\text{r}\right)}}{d_{3 \mathrm{D}}}}_{\text {LoS signal }}\\
& +\underbrace{\frac{\Gamma\left(\theta_\text{ref}\right) \sqrt{\mathrm{G}_{\mathrm{gs}}\left(\phi_\text{t1}, \theta_\text{t1}\right) \mathrm{G}_{\mathrm{uav}}\left(\phi_\text{r1}, \theta_\text{r1}\right)} e^{-j \Delta \tau}}{d_\mathrm{1}+d_\mathrm{2}}}_{\text {ground reflected signal }}\biggr\rvert^2,
\end{aligned}
\label{eqn:two-ray-model}
\end{equation}
where \(\mathrm{A}_{\text {trpl}}\) represents the signal attenuation factor, defined as the ratio of the received power to the transmitted power. The terms $G_{\text{gs}}(\cdot)$ and $G_{\text{uav}}(\cdot)$ denote the antenna gain of the GS and UAV antennas, with their corresponding elevation and azimuth angles. The four elevation angles $\theta_\text{t}$, $\theta_\text{t1}$, $\theta_\text{r}$, and $\theta_\text{r1}$ are demonstrated in Fig.~\ref{fig:diagram_two-ray-model}. Additionally, $l_{\text{wave}}$ is the wavelength, $\Delta \tau$ is the phase delay between two paths, and $\Gamma\left(\theta_\text{ref}\right)$ is the ground reflection coefficient. The corresponding path loss, typically expressed on a dB scale, is given by:
\begin{equation}
\mathrm{PL}_{\text{trpl}}^{\text{(dB)}} = -10 \, \log_{10} \left( \mathrm{A}_{\text{trpl}} \right)~.
\label{eqn:two_ray_pl}
\end{equation}
The estimated received power using TRPL, denoted as \(\hat{r}_{\mathrm{trpl}}\), is given by:
\begin{equation}
\label{eqn:received_power_tworay}
\hat{r}_{\mathrm{trpl}}^{(\text{dB})} = \mathrm{P}_{\mathrm{Tx}}^{(\text{dB})} - \mathrm{PL}_{\mathrm{trpl}}^{\text{(dB)}},
\end{equation}
where \(\mathrm{P}_{\mathrm{Tx}}^{(\text{dB})}\) is the transmit power in dB and \(\mathrm{PL}_{\mathrm{trpl}}^{\text{(dB)}}\) is the TRPL-estimated deterministic path loss given in~\eqref{eqn:two_ray_pl}.

\subsection{Spatial Correlation of Shadow Fading (SF)}
\label{sec:sf_property}
\begin{figure}% 
    \centering
		\vspace{-0mm}
    \includegraphics[width=\linewidth]{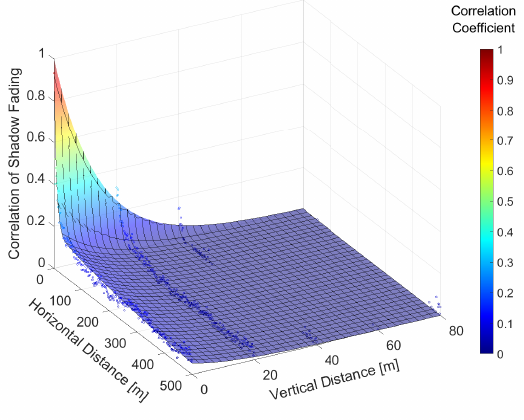}
    \hspace{4mm}
    \vspace{-3mm}
    \caption{Visualization of the spatial correlation of shadow fading (SF), \(w\), as a function of the horizontal and vertical distances, \(d_{\mathrm{h}}\) and \(d_{\mathrm{v}}\).}
    \vspace{0mm}
    \label{fig:correlation-model}
\end{figure}

Unlike deterministic path loss, SF follows a stochastic process characterized by spatial correlation across 3D space. This phenomenon, which captures large-scale fading caused by the presence of significant physical obstacles, varies smoothly throughout the 3D environment \cite{szyszkowicz2010feasibility}. Let $w$ represent the SF component at a given location. For any two 3D coordinates, $l_i$ and $l_j$, let $d_{\mathrm{h}}$ and $d_{\mathrm{v}}$ denote their respective horizontal and vertical separation distances. The spatial correlation of the SF between $l_i$ and $l_j$, denoted by $R_w(l_i, l_j)$, is expressed as:
\begin{equation}
R_w\left(l_i, l_j\right)=\frac{\mathbb{E}\left[w(l_i) w(l_j)\right]}{\sigma_w^2},%=R(d_{\mathrm{v}}, d_{\mathrm{h}}),
\label{eqn:corr_definition}
\end{equation}
where $w(\cdot)$ represents the SF at the corresponding location and $\sigma_w^2$ denotes the variance of $w$. The spatial correlation $R_w(l_i, l_j)$ is characterized by exponential and biexponential distributions relative to the vertical distance $d_{\mathrm{v}}$ and horizontal distance $d_{\mathrm{h}}$, respectively, as formulated in \cite{maeng2023kriging, rahman20243d}:
\begin{equation}
\label{eqn:corr_model}
R_w\left(l_i, l_j\right)=e^{-qd_{\mathrm{v}}}[ae^{-p_1d_{\mathrm{h}}}+(1-a)e^{-p_2d_{\mathrm{h}}}],
\end{equation}
where $a$, $p_1$, $p_2$, and $q$ are constants that can be estimated from field measurement data. An illustration of \(R_w\left(l_i, l_j\right)\) with respect to \(d_{\mathrm{v}}\) and \(d_{\mathrm{h}}\) is provided in Fig.~\ref{fig:correlation-model}.

\subsection{Problem Setup}
\label{sec:pro_set}
This study addresses 3D REM reconstruction by estimating independent 2D slices at discrete UAV measurement altitudes from limited sparse samples. The problem is formulated as follows: for a given measurement campaign $\mathcal{E}$ conducted at a constant altitude $H$, we have a set of $n$ observed signal power data points. These measurements are denoted as $\mathrm{P}_{\mathrm{Rx}}^{\text{(dB)}}(l_i)$, where $i \in \{1, \dots, n\}$, and $l_i$ represents the 3D coordinate of the $i$-th measurement location. To simulate a \textbf{highly undersampled regime}, a small subset of $\mathrm{M}$ points is randomly selected from the $n$ measurements to serve as the sparse samples for REM reconstruction. We vary $\mathrm{M}$ from $10$ to $250$ to analyze performance across various levels of data sparsity. The remaining $(n - \mathrm{M})$ points are reserved as a ground-truth testing set to evaluate reconstruction accuracy at unsampled locations.

Furthermore, the transmitter location, the anechoic chamber antenna patterns of the isolated antennas, and the transmit power are provided. To facilitate parameter learning, a separate UAV campaign $\mathcal{E}'$, conducted at an altitude $H'$ with $n'$ measurements ($n' > \mathrm{M}$), is utilized. This auxiliary dataset serves as the \textbf{training dataset} for estimating environmental spatial correlation parameters in~\eqref{eqn:corr_model} or for optimizing the weights of deep learning models, whereas the limited $\mathrm{M}$ samples from the target experiment $\mathcal{E}$ are used during inference. To ensure a rigorous and valid train-test split, we maintain strict spatial separation between experiments $\mathcal{E}$ and $\mathcal{E}'$. This is achieved by either ensuring a vertical separation ($|H - H'| \geq 20$~m) or by verifying that the 2D horizontal trajectories of $\mathcal{E}$ and $\mathcal{E}'$ are sufficiently disjoint when conducted at the same altitude.

\section{UAV Measurement Datasets}
\label{sec:datasets}
This section first details the hardware configurations and experimental environments of the empirical UAV measurement datasets used in this study, and subsequently introduces each individual dataset along with its technical specifications.
\subsection{Environment and Hardware Setup}
\begin{figure}[!t]
    \centering
    \begin{minipage}{0.48\linewidth}
        \centering
        \includegraphics[width=\linewidth]{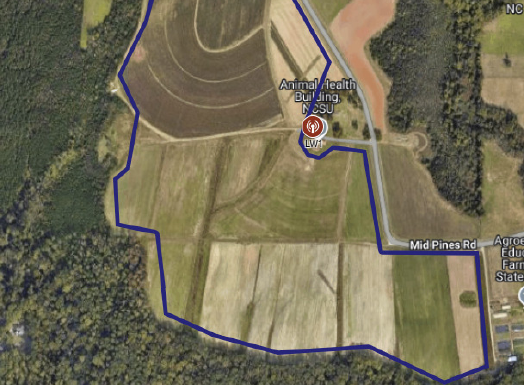}
        \subcaption{Site Map}%\label{fig:sub1}
    \end{minipage}%
    \hfill
    \begin{minipage}{0.48\linewidth}
        \centering
        \includegraphics[width=\linewidth]{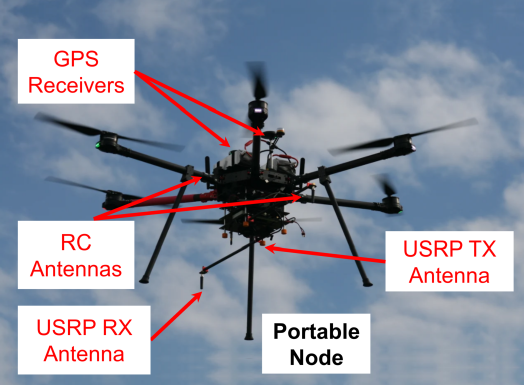}
        \subcaption{UAV-Based Spectrum Sensor}\label{fig:sub2}
    \end{minipage}
    
    \vspace{0.2cm}

    \begin{minipage}{0.48\linewidth}
        \centering
        \includegraphics[width=\linewidth]{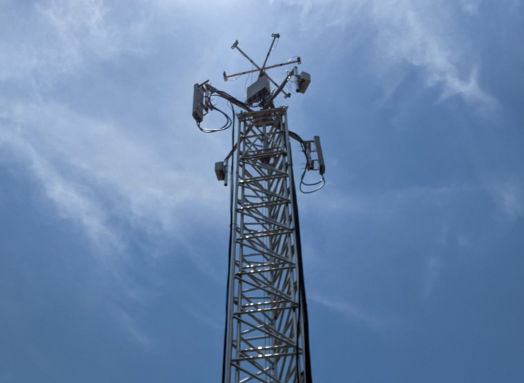}
        \subcaption{BS}\label{fig:sub3}
    \end{minipage}%
    \hfill
    \begin{minipage}{0.48\linewidth}
        \centering
        \includegraphics[width=\linewidth]{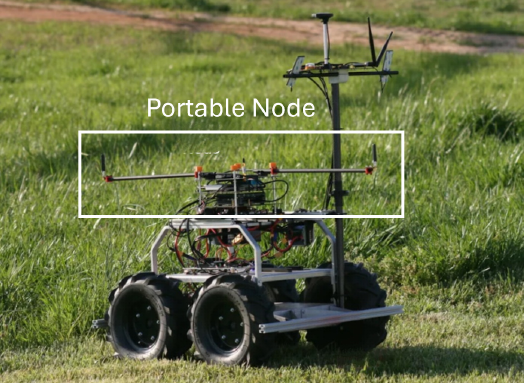}
        \subcaption{UGV}\label{fig:sub4}
    \end{minipage}
    
    \caption{Environment and communication nodes in the data collection setup: (a) Google Earth view of the site, (b) UAV-based measurement system, and (c) and (d) ground nodes.}
    \label{fig:aerpaw_site}
    \vspace{-4mm}
\end{figure}

In this study, we utilize three publicly available UAV-based RSRP measurement datasets~\cite{IEEEDataPort_2, masrur2025collection, IEEEDataPort_4}, collected at one of the Aerial Experimentation and Research Platform on Advanced Wireless~(AERPAW)~\cite{9061157,gurses2025digital} experimentation sites located at the Lake Wheeler Field Labs in Raleigh, North Carolina, USA. A geographical map of the deployment environment is illustrated in Fig.~\ref{fig:aerpaw_site}(a). The UAV is equipped with an AERPAW portable node (APN), which integrates a Universal Software Radio Peripheral (USRP) and specialized RF frontends to function as a mobile spectrum sensor. Fig.~\ref{fig:aerpaw_site}(b) displays the UAV with the mounted APN, highlighting its core components. Specifically, a wideband antenna (SA-1400-5900)~\cite{sa_1400} is mounted on a horizontal rod beneath the UAV chassis. Across the datasets, two distinct GS setups are employed: fixed BSs equipped with RM-WB1-DN antennas~\cite{mobilemarkrmwb1}, as shown in Fig.~\ref{fig:aerpaw_site}(c), and mobile APNs mounted on Uncrewed Ground Vehicles~(UGV). The APNs utilize the same SA-1400-5900 antenna configuration, as depicted in Fig.~\ref{fig:aerpaw_site}(d).

\subsection{Individual Datasets and Experiments}
Table~\ref{tab:exp_list} provides a short designation for all experiments used in this study, along with their respective properties.
\subsubsection{AERPAW LTE I/Q measurement dataset~\cite{IEEEDataPort_2}}
\begin{figure}[!t]
\centerline{\includegraphics[width=\linewidth]{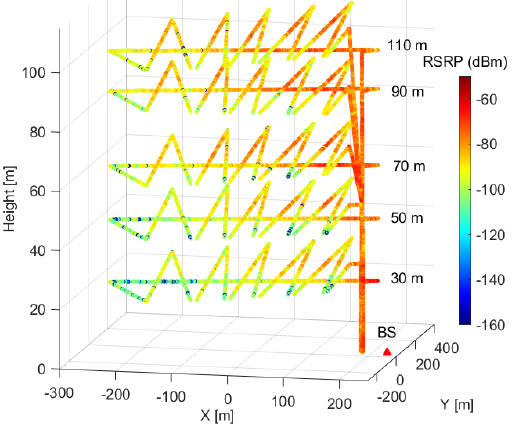}}
\caption{AERPAW LTE I/Q measurement dataset~\cite{IEEEDataPort_2}.}
\label{fig:uav_trajectory}
\vspace{-2mm}
\end{figure}

A BS acted as the signal source, while the UAV collected I/Q samples at five different altitudes ranging from \(30\)~m to \(110\)~m, following a zigzag trajectory at each altitude. The resulting RSRP map is presented in Fig.~\ref{fig:uav_trajectory}.

\subsubsection{AERPAW Find-a-Rover (AFAR) dataset~\cite{masrur2025collection}}
\begin{figure*}[htbp]
    \begin{subfigure}{0.19\textwidth}
        \centering
        \includegraphics[width=\linewidth]{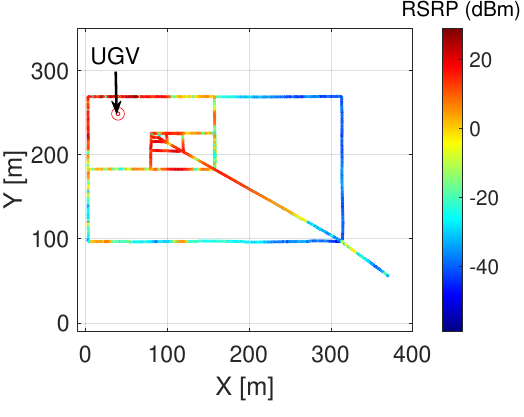}
        \caption{B1 (21 m)}
        %\label{fig:sub1}
    \end{subfigure}
    \hfill
    \begin{subfigure}{0.19\textwidth}
        \centering
        \includegraphics[width=\linewidth]{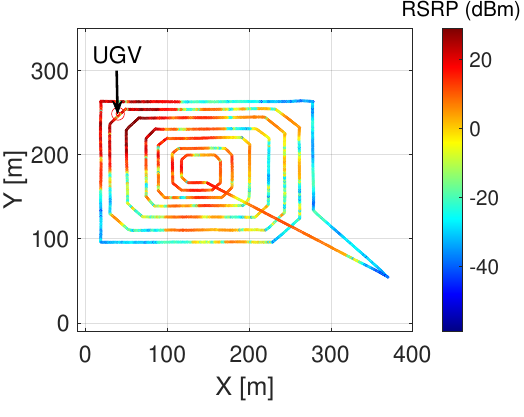}
        \caption{B2 (28 m)}
        %\label{fig:sub1}
    \end{subfigure}
    \hfill
    \begin{subfigure}{0.19\textwidth}
        \centering
        \includegraphics[width=\linewidth]{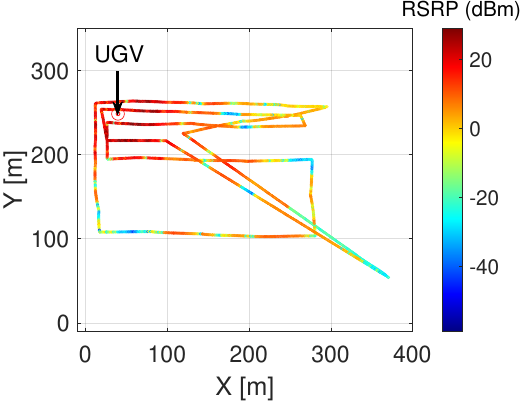}
        \caption{B3 (28 m)}
        %\label{fig:sub1}
    \end{subfigure}
    \hfill
    \begin{subfigure}{0.19\textwidth}
        \centering
        \includegraphics[width=\linewidth]{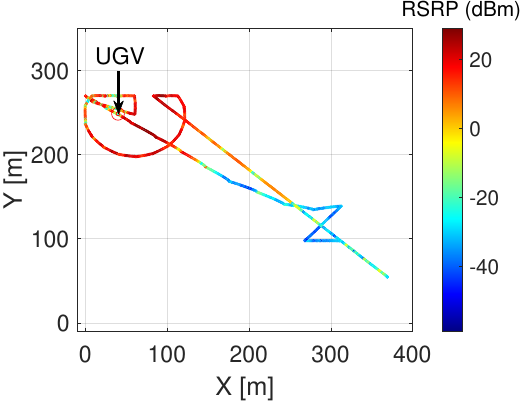}
        \caption{B4 (37 m)}
        %\label{fig:sub1}
    \end{subfigure}
    \hfill
    \begin{subfigure}{0.19\textwidth}
        \centering
        \includegraphics[width=\linewidth]{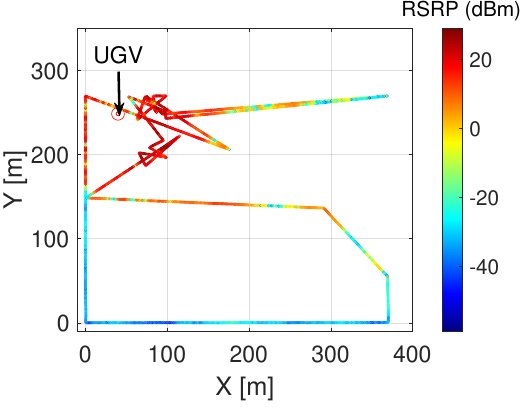}
        \caption{B5 (47 m)}
        %\label{fig:sub1}
    \end{subfigure}
    \hfill
        \caption{AERPAW Find-a-Rover (AFAR) dataset~\cite{IEEEDataPort_3}, illustrating the UGV location and the UAV trajectories. The subfigure captions denote the specific experiment labels and their respective flight altitudes.}
    \label{fig:trajectories_afar}
\end{figure*}

The GS was a UGV with an antenna height of \(1.5\)~m above the ground. Across the five experimental campaigns in this dataset, the UAV flight altitudes ranged from $28$~m to $47$~m. The UAV trajectories were highly diverse, as illustrated in Fig.~\ref{fig:trajectories_afar}.

\subsubsection{AERPAW multi-band dataset~\cite{IEEEDataPort_4}}
\begin{figure}[]
\centerline{\includegraphics[width=\linewidth]{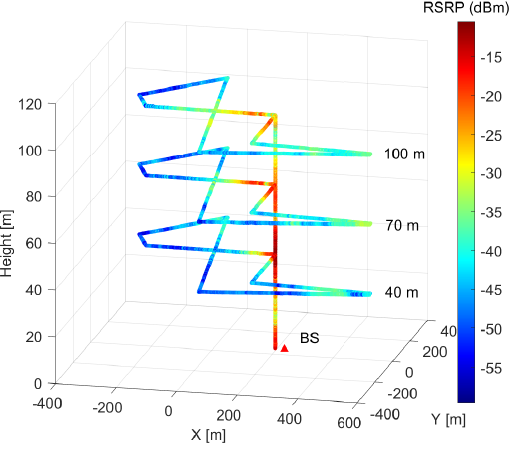}}
\vspace{4mm}
\caption{Illustration of measured RSRPs at $5$~MHz bandwidth for AERPAW multi-band dataset~\cite{IEEEDataPort_4}.}
\label{fig:uav_trajectory_cole}
\end{figure}

The GS was the AERPAW BS, consistent with the AERPAW LTE dataset~\cite{IEEEDataPort_2}. Data collection was conducted at UAV altitudes of $40$~m, $70$~m, and $100$~m, with measurement bandwidths ranging from $1.25$~MHz to $5$~MHz. The resulting RSRP map for a representative bandwidth configuration is illustrated in Fig.~\ref{fig:uav_trajectory_cole}.
\begin{table}[!t]
\centering
\renewcommand{\arraystretch}{1.3} % Increase row height by 1.5 times
\caption{Experiments considered in this study.}
\begin{tabular}{cccccc}
\hline
\textbf{Dataset} & \textbf{Label} & \textbf{Altitude} & \textbf{Bandwidth} & \textbf{Carrier}  & \textbf{GS}                                \\ \hline
LTE I/Q & A1                         & 30~m               & 1.4~MHz             & 3.5~MHz & BS \\ \hline
LTE I/Q & A2                         & 50~m               & 1.4~MHz             & 3.5~MHz & BS \\ \hline   
LTE I/Q & A3                         & 70~m               & 1.4~MHz             & 3.5~MHz & BS \\ \hline 
LTE I/Q & A4                         & 90~m               & 1.4~MHz             & 3.5~MHz & BS \\ \hline 
LTE I/Q & A5                         & 110~m               & 1.4~MHz             & 3.5~MHz & BS \\ \hline \hline
AFAR & B1                         & 21~m               & 125~kHz             & 3.32~MHz & UGV \\ \hline
AFAR & B2                         & 28~m               & 125~kHz             & 3.32~MHz & UGV \\ \hline   
AFAR & B3                         & 28~m               & 125~kHz             & 3.32~MHz & UGV \\ \hline 
AFAR & B4                         & 37~m               & 125~kHz           & 3.32~MHz & UGV \\ \hline 
AFAR & B5                         & 47~m               & 125~kHz              & 3.32~MHz & UGV \\ \hline \hline
Multi-band & C1                         & 40~m               & 5~MHz             & 3.3~MHz & BS \\ \hline
Multi-band & C2                         & 70~m               & 5~MHz             & 3.3~MHz & BS \\ \hline   
Multi-band & C3                         & 100~m               & 5~MHz             & 3.3~MHz & BS \\ \hline 
Multi-band & C4                         & 40~m               & 1.25~MHz             & 3.3~MHz & BS \\ \hline 
Multi-band & C5                         & 40~m               & 2.5~MHz             & 3.3~MHz & BS \\ \hline
\end{tabular}
\label{tab:exp_list}
\end{table}

\section{3D REM Reconstruction Algorithm} \label{sec:sensing_method}
In this section, we present the 3D REM reconstruction framework, which integrates a deterministic radio propagation model for large-scale path loss prediction with a data-driven approach for stochastic component estimation. The reconstructed received power at an unsampled location, $\hat{r}^{\text{(dB)}}$, is expressed on the dB scale as:
\begin{equation}
\hat{r}^{\text{(dB)}} = \hat{r}_{\mathrm{trpl}}^{\text{(dB)}} + \hat{w},
\label{eqn:predicted_power}
\end{equation}
where $\hat{r}_{\mathrm{trpl}}^{\text{(dB)}}$ represents the predicted received power from the TRPL model, calculated via~\eqref{eqn:received_power_tworay}, and $\hat{w}$ is the data-driven SF prediction detailed hereafter.

The data-driven SF prediction techniques utilize a subset of $\mathrm{M}$ RSRP observations from $\mathcal{E}$. By subtracting the TRPL-based received power from these observations, we obtain the $\mathrm{M}$ residual SF components, denoted as $w_{\mathrm{trpl}}$, as follows:
\begin{equation}
w_{\mathrm{trpl}} = \mathrm{P}_{\mathrm{Rx}}^{\text{(dB)}} - \hat{r}_{\mathrm{trpl}}^{\text{(dB)}} = Z,
\label{subeq:z1}
\end{equation}
where $\mathrm{P}_{\mathrm{Rx}}^{\text{(dB)}}$ is the measured RSRP sample, and $Z$ is used interchangeably with $w$ (e.g., in \eqref{eqn:corr_definition}) to denote the stochastic SF component. These known $Z$ values serve as the primary inputs for predicting the SF at unsampled locations by exploiting its spatial correlation properties.

This study introduces a novel algorithm that leverages both GPR and MC for SF prediction. For performance comparison, we employ OK~\cite{wackernagel2003ordinary}, SK~\cite{olea1999simple}, and their respective trans-Gaussian variants~\cite{cressie2015statistics}. In the following subsections, we provide the formulations for OK, SK, trans-Gaussian Kriging, and GPR, followed by our proposed MC-assisted GPR, which accounts for non-uniform spatial smoothness across regions. It reconstructs the map in two layers: one dedicated to local deep-shadowed zones and another for the remaining regions.

\subsection{Ordinary Kriging (OK)}
\label{subsection:ordinary_Kriging}
Let the 3D coordinates of the $\mathrm{M}$ sparse measurement locations be denoted by $\{l_1, \dots, l_\mathrm{M}\}$, where $\mathrm{M}$ represents the number of available sparse samples. The corresponding SF components at these locations are denoted by $Z(l_i)$, where $l_i = (\psi_i, \zeta_i, h_i)$ corresponds to the latitude, longitude, and altitude, respectively.
OK employs a linear combination of these known measurements to predict the signal $\hat{Z}(l_0)$ at a new location $l_0$. Its objective is to minimize the squared estimation error while ensuring that the sum of the weights used for the linear combination equals 1. Formally, it can be written as:
\begin{equation}
\label{eqn:ordinary_kriging_inference}
\begin{aligned}
\min _{w_{1}, \ldots, w_{\mathrm{M}}} & \mathbb{E}\left[\left(\hat{Z}\left(l_0\right)-Z\left(l_0\right)\right)^2\right], \\
\text { s.t. } & \hat{Z}\left(l_0\right)=\sum_{i=1}^\mathrm{M}  w_{i} Z(l_i), \\
& \sum_{i=1}^\mathrm{M} w_{i}=1.
\end{aligned}
\end{equation}
This constrained minimization problem can be solved using the method of Lagrange multipliers, which reduces the optimization to the following system of linear equations:
\begin{equation}
\small
\begin{aligned}
& {\left[\begin{array}{cccc}
\mathbf{\gamma}(l_1,l_1) & \cdots & \mathbf{\gamma}\left(l_1,l_\mathrm{M}\right) & 1 \\
\vdots & \ddots & \vdots & \vdots \\
\mathbf{\gamma}(l_\mathrm{M},l_1) & \cdots & \mathbf{\gamma}\left(l_\mathrm{M},l_\mathrm{M}\right) & 1 \\
1 & \cdots & 1 & 0
\end{array}\right]\left[\begin{array}{c}
w_{1} \\
\vdots \\
w_{\mathrm{M}} \\
\mu
\end{array}\right]} =\left[\begin{array}{c}
\mathbf{\gamma}\left(l_0,l_1\right) \\
\vdots \\
\mathbf{\gamma}\left(l_0,l_\mathrm{M}\right) \\
1
\end{array}\right] ,
\end{aligned}
\label{eq:ordinary_kirging_soln}
\end{equation}
where $\mu$ is the Lagrange multiplier, and $\mathbf{\gamma}\left(l_i,l_j\right)$ denotes the semivariogram~\cite{olea1994fundamentals} between two locations $l_i$ and $l_j$ with respect to $Z$, defined as:
\begin{equation}
\mathbf{\gamma}\left(l_i,l_j\right)  = \mathrm{Var}\big(Z(l_i)-Z(l_j)\big),
\end{equation}
where \(\mathrm{Var(\cdot)}\) denotes the variance operator. For zero-mean SF \(Z\), the semivariogram \(\mathbf{\gamma}\left(l_i,l_j\right)\) is related to spatial correlation \(R_z\left(l_i, l_j\right)\) by the following expression:
\begin{equation}
\mathbf{\gamma}\left(l_i,l_j\right) = \sigma^2_Z \left[ 1- R_z\left(l_i, l_j\right)\right],
\end{equation}
where $\sigma_Z^2$ is the variance of \(Z\). The analytical expression for the mean squared error (MSE) in predicting at $l_0$ is given by:
\begin{equation}
\label{eqn:ordianry_Kriging_variance}
\sigma_{\text{ok}}^2\left(l_0\right) = \sum_{i=1}^\mathrm{M} w_i \gamma\left(l_0,l_i\right)+\mu.
\end{equation}
It should be noted that OK is classified as a Best Linear Unbiased Estimator (BLUE), where the unbiasedness is ensured by assuming a locally constant but unknown mean and imposing the constraint that the weighting parameters sum to unity.

\subsection{Simple Kriging (SK)}
Given \(\mathrm{M}\) sparse measurement samples, SK estimates the unknown value at a target location $l_0$, \(\hat{Z}\left(l_0\right)\), through a linear combination of the observed samples ${Z(l_1), ..., Z(l_\mathrm{M})}$. Unlike OK, SK does not impose a constraint on the sum of the weighting parameters. A constant mean $m_Z$ is assumed for the process $Z$, which is typically around zero for SF. The objective of SK is to minimize the MSE at an arbitrary location $l_0$, as formulated below: 
\begin{eqnarray}
\label{eqn:simple_kriging_inference}
\min _{w_{1}, \ldots, w_{\mathrm{M}}} & \mathbb{E}\left[\left(\hat{Z}\left(l_0\right)-Z\left(l_0\right)\right)^2\right], \nonumber \\
\text { s.t. } & \hat{Z}\left(l_0\right)=m_Z + \sum_{i=1}^\mathrm{M}  w_{i} \left(Z\left(l_i\right) - m_Z\right).
\end{eqnarray}
The optimum weights for SK can be computed by solving the following set of linear equations:
\begin{equation}
\small
\label{eqn:simple_Kriging_soln}
\begin{aligned}
& {\left[\begin{array}{cccc}
C(l_1,l_1) & \cdots & C\left(l_1,l_\mathrm{M}\right) \\
\vdots & \ddots & \vdots \\
C(l_\mathrm{M},l_1) & \cdots & C\left(l_\mathrm{M},l_\mathrm{M}\right)
\end{array}\right]\left[\begin{array}{c}
w_{1} \\
\vdots \\
w_{\mathrm{M}} \\
\end{array}\right]} =\left[\begin{array}{c}
C\left(l_0,l_1\right) \\
\vdots \\
C\left(l_0,l_\mathrm{M}\right)
\end{array}\right] ,
\end{aligned}
\end{equation}
where $C(l_i, l_j)$ represents the spatial covariance of $Z$ between two locations $l_i$ and $l_j$, and is directly related to the spatial correlation $R_Z(l_i, l_j)$ through the SF variance $\sigma^2_Z$ as follows:
\begin{equation}
C(l_i, l_j) = \sigma^2_Z R_z\left(l_i, l_j\right).
\end{equation}
The prediction MSE for SK is given as follows:
\begin{equation}
\label{eqn:simple_Kriging_variance}
\sigma_\text{sk}^2\left(l_0\right) = \sigma_Z^2 - \sum_{i=1}^\mathrm{M} w_i C\left(l_0,l_i\right).
\end{equation}
SK is recognized as the Best Linear Predictor~(BLP) and is inherently unbiased due to the assumption of a known, constant mean. Furthermore, if the process $Z$ is Gaussian, SK also serves as the global best predictor among all (linear and nonlinear) estimators.

\subsection{Trans-Gaussian Kriging} 
In some cases, the SF data~\(Z\) exhibit a skewed distribution. We know that the Gaussian distribution ensures the optimality of a linear predictor over all possible predictors, including nonlinear ones, in terms of minimizing MSE~\cite[Chapter 3]{cressie2015statistics}. Hence, we can transform $Z$ into $U$ that has a standard normal distribution $\mathcal{N}(0,1)$. The transformation function $f(\cdot)$ can be obtained as follows~\cite{casella2021statistical}:
\begin{equation}
U(l) = \operatorname{CDF}_{\mathcal{N}(0,1)}^{-1}\Big(\operatorname{CDF}_{Z}\big(Z(l)\big)\Big) \equiv f\big(Z(l)\big),
\end{equation}
where $\operatorname{CDF}(\cdot)$ is the cumulative distribution function of a random variable.
Let the mean of $U$ be denoted as $m_U$ and the variance of $U$ as $\sigma^2_U$, the inverse function of $f(\cdot)$ as $\phi(\cdot)$, the OK prediction of $U$ as $\hat{U}_\text{ok}(\cdot)$, and the SK prediction of $U$ as $\hat{U}_\text{sk}(\cdot)$. Their corresponding MSE, calculated using~\eqref{eqn:ordianry_Kriging_variance}~and~\eqref{eqn:simple_Kriging_variance}, are denoted as $\sigma_{U,\text{ok}}^2\left(\cdot\right)$ and $\sigma_{U,\text{sk}}^2\left(\cdot\right)$, respectively, and the Lagrange multiplier in \eqref{eq:ordinary_kirging_soln} is $\mu_U$.

After predicting variable $U$ at location~$l_0$, we need to transform it back into the $Z$ domain using~$\phi(\cdot)$. Moreover, it turns out that estimation in the $Z$ domain introduces a prediction bias if the second-order derivative of $\phi(\cdot)$ computed at $m_U$ is nonzero. To compensate for this bias, the approximately unbiased solution for OK is given as follows~\cite{cressie2015statistics}:
\begin{equation}
\hat{Z}_\text{ok}\!\left(l_0\right)=\phi\!\left(\hat{U}_\text{ok}(l_0)\right)+\phi^{\prime \prime}\!\left(m_U\right)\left[\sigma_{U, \text{ok}}^2\left(l_0\right) \big/ 2\!-\!\mu_U\right].
\end{equation}
The equation above is derived using the second-order delta method.
Similarly, for SK, the approximately unbiased solution is provided by~\cite{cressie2015statistics}:
\begin{equation}
\small
\begin{aligned}
\hat{Z}_\text{sk}\!\left(l_0\right)&=\phi\left(\hat{U}_\text{sk}(l_0)\right) +\frac{\phi^{\prime \prime}\!\left(m_U\right)} {2}\Big[\sigma_{U,\text{sk}}^2\left(\mathbf{s}_0\right)-\sum_{i=1}^\mathrm{M} w_i C\left(l_0,l_i\right)\Big].
\end{aligned}
\end{equation}

\subsection{Gaussian Process Regression (GPR)}
In contrast to SK, which models $Z$ as a direct random process with variance $\sigma_Z^2$, GPR treats $Z$ as the sum of a latent stochastic process $Y$ and an additive measurement noise component $\delta_m$. Specifically, $Z(l) = Y(l) + \delta_m$, where $l$ denotes the measurement location and $\delta_m \sim \mathcal{N}(0, \sigma_{\text{GP}}^2)$ represents Gaussian noise with variance $\sigma_{\text{GP}}^2$. The latent process $Y$ is assumed to have a zero mean ($m_Y = 0$), with a spatial covariance governed by $C_Y(l_i, l_j) = \sigma_Y^2 R_Y(l_i, l_j)$, where $\sigma_Y^2$ is the signal variance and $R_Y(\cdot)$ is the correlation function. The covariance of the noisy observations $Z$ between two locations $l_i$ and $l_j$ then can be expressed as follows:
\begin{equation}
C_Z(l_i, l_j) = C_Y(l_i, l_j) + \sigma_{\text{GP}}^2 \delta_{ij},
\end{equation}
where $\delta_{ij}$ is the Kronecker delta, which is equal to 1 if $i=j$ and 0 otherwise. For a zero-mean process, the GPR estimate $\hat{Z}(l_0)$ at an unobserved location $l_0$ is given by:
\begin{equation}
\hat{Z}(l_0) = \sum_{i=1}^\mathrm{M} w_i Z(l_i).
\label{eqn:gpr_pred}
\end{equation}

The optimal weights $w_i$ for GPR are computed by solving the following system of linear equations, which incorporates the noise regularization along the diagonal:
\begin{equation}
\label{eqn:gpr_soln}
\begin{aligned}
& {\left[\begin{array}{cccc}
C_Y(l_1,l_1) + \sigma_{\text{GP}}^2 & \cdots & C_Y\left(l_1,l_\mathrm{M}\right) \\
\vdots & \ddots & \vdots \\
C_Y(l_\mathrm{M},l_1) & \cdots & C_Y\left(l_\mathrm{M},l_\mathrm{M}\right) + \sigma_{\text{GP}}^2
\end{array}\right]\left[\begin{array}{c}
w_{1} \\
\vdots \\
w_{\mathrm{M}} \\
\end{array}\right]}\\
&=\left[\begin{array}{c}
C_Y\left(l_0,l_1\right) \\
\vdots \\
C_Y\left(l_0,l_\mathrm{M}\right)
\end{array}\right] .
\end{aligned}
\end{equation}
The prediction variance for GPR, which represents the uncertainty of reconstruction at location $l_0$, is given as follows:
\begin{equation}
\sigma_{\text{GPR}}^2(l_0) = \sigma_Y^2 + \sigma_{\text{GP}}^2 - \sum_{i=1}^\mathrm{M} w_i C_Y(l_0, l_i).
\label{eqn:gpr_var}
\end{equation}
While the expressions in \eqref{eqn:gpr_soln} and \eqref{eqn:gpr_var} are structurally similar to those of SK, GPR is regularized by the noise variance $\sigma_{\text{GP}}^2$, making the reconstruction more robust to noise.

\subsection{MC-Assisted GPR for Deep-Shadow Extraction}
\label{subsec:mat_completion}
The core principle of the proposed MC-assisted GPR framework lies in decomposing the REM into two distinct components: a smooth spatial trend and a highly varying local component representing the deep-shadowed regions. The MC algorithm is first employed to estimate the underlying smooth trend. Subsequently, the deep-shadow regions are identified and enhanced via a grayscale dilation operation. The methodology is detailed as follows:
\begin{algorithm}[!t]
    \caption{MC-Assisted GPR.}
    \label{alg:mc_algo}
        \begin{algorithmic}[1]
            \Procedure{$\textnormal{Main}$}{}
            \State $\textbf{Input:} \boldsymbol{D}_{\text{Z}} \in \mathbb{R}^{\mathrm{M}}, \boldsymbol{D}_{\psi} \in \mathbb{R}^{\mathrm{M}}, \boldsymbol{D}_{\zeta} \in \mathbb{R}^{\mathrm{M}}, \psi_{\text{0}} \in \mathbb{R}, \zeta_{\text{0}} \in \mathbb{R}, T_\text{v} \in \mathbb{R}$
            \State $\textbf{REM Decomposition:}$
            \State $\hat{\boldsymbol{Z}}_{\text{mc}},\boldsymbol{Z_{\text{}}} \gets \,\text{MatrixCompletion}(\boldsymbol{D}_{\text{Z}}, \boldsymbol{D}_{\psi}, \boldsymbol{D}_{\zeta})$
            \State $\hat{\boldsymbol{Z}}_{\Delta} \gets \,\boldsymbol{Z_{\text{}}}-\hat{\boldsymbol{Z}}_{\text{mc}}$
            \State $\hat{\boldsymbol{Z}}_{\text{ds}} \gets \,\boldsymbol{Z}_{\Delta}\,\cdot\,\left(\,\abs{\boldsymbol{Z}_{\Delta}}>T_\text{v}\right)$
            \State $\hat{\boldsymbol{Z}}_{\text{smooth}} \gets \,\boldsymbol{Z_{\text{}}}-\hat{\boldsymbol{Z}}_{\text{ds}}$
            \State \textbf{Deep-Shadow Enhancement:}
            \State $\hat{\boldsymbol{Z}}_{\text{ds}} \gets \text{GrayscaleDilation}(\hat{\boldsymbol{Z}}_{\text{ds}})$
            \State \textbf{Prediction:}
            \State $\hat{Z}_{\text{}} \gets \text{interp2}(\hat{\boldsymbol{Z}}_{\text{smooth}}+\hat{\boldsymbol{Z}}_{\text{ds}}, \psi_\text{0}, \zeta_\text{0},``\text{spline}")$
            \EndProcedure
            
            \Procedure{$\textnormal{MatrixCompletion}$}{}%\Comment{The 
            \State $\textbf{Input:} \boldsymbol{D}_{\text{Z}} \in \mathbb{R}^{\mathrm{M}}, \boldsymbol{D}_{\psi} \in \mathbb{R}^{\mathrm{M}}, \boldsymbol{D}_{\zeta} \in \mathbb{R}^{\mathrm{M}}, \alpha\in \mathbb{R}$
            \State $\textbf{Initialization:}$
            \State  $\text{d}_{\text{grid}} \gets $ Constant, equivalent to 5\,m
            \State  $\boldsymbol{G}_{\psi} \gets \text{min}(\boldsymbol{D}_{\psi}):\text{d}_{\text{grid}}:\text{max}(\boldsymbol{D}_{\psi})$
            \State  $\boldsymbol{G}_{\zeta} \gets \text{min}(\boldsymbol{D}_{\zeta}):\text{d}_{\text{grid}}:\text{max}(\boldsymbol{D}_{\zeta})$
            \State $N_\text{r} = \text{count}(\boldsymbol{G}_{\psi})$, $N_\text{c} = \text{count}(\boldsymbol{G}_{\zeta})$
            \State  $\boldsymbol{S}_{i} \gets \{1,2,\dots,N_{\text{r}}\}$, $\boldsymbol{S}_{j} \gets \{1,2,\dots,N_{\text{c}}\}$ 
            \State $\boldsymbol{Z_{\text{}}} \in \mathbb{R}^{N_{\text{r}} \times N_{\text{c}}}$ \Comment{GPR Prediction}
            \State $\boldsymbol{\Sigma_{\text{}}} \in \mathbb{R}^{N_{\text{r}} \times N_{\text{c}}}$ \Comment{Prediction Variance}
            \State \textbf{Local interpolation:}
            \For{$i \in \boldsymbol{S}_{i}$}
            \For{$j \in \boldsymbol{S}_{j}$}
            \State $z_{\text{k}}, v_{\text{k}} \gets\,\text{GPR}(\boldsymbol{D}_{\text{Z}}, \boldsymbol{D}_{\psi}, \boldsymbol{D}_{\zeta}, \boldsymbol{G}_{\psi}(i), \boldsymbol{G}_{\zeta}(j))$
            \State $\boldsymbol{Z_{\text{}}}_{ij}, \boldsymbol{\Sigma_{\text{}}}_{ij} \gets z_{\text{k}}, v_{\text{k}}$
            \EndFor
            \EndFor
            \State \textbf{Global matrix completion:}
            \State $\hat{\boldsymbol{Z}}_{\text{mc}} \gets \text{NuclearNormMin}(\boldsymbol{Z_{\text{}}},\boldsymbol{\Sigma_{\text{}}}, \alpha)$\Comment{Algorithm~\ref{alg:mc_func}}
            \State $\textbf{Output:}\,\hat{\boldsymbol{Z}}_{\text{mc}},\boldsymbol{Z_{\text{}}}$
            \EndProcedure
        \end{algorithmic}
\end{algorithm}

\subsubsection{Sparse Measurements Into Matrix}
MC is a reconstruction process that takes a partially observed matrix as input and recovers the full matrix as output. Since this method is strictly applicable to grid-structured data, the sparse observations must first be transformed into a matrix representation. To facilitate this, we define a bounded 2D region in the horizontal plane that encapsulates the entire UAV trajectory. This region is then discretized into $5$~m $\times$~$5$~m square grids, where each grid cell corresponds to a unique entry in the matrix. Given $\mathrm{M}$ observation samples of $Z$ (defined by latitudes $\boldsymbol{D}_{\psi} \in \mathbb{R}^{\mathrm{M}}$, longitudes $\boldsymbol{D}_{\zeta} \in \mathbb{R}^{\mathrm{M}}$, and SF values $\boldsymbol{D}_{\text{Z}} \in \mathbb{R}^{\mathrm{M}}$), we construct an observation matrix $\boldsymbol{Z}$ by estimating $Z$ at each grid center using~\eqref{eqn:gpr_pred}, denoted as $\boldsymbol{Z}_{ij}$. Concurrently, we form a standard deviation matrix $\mathbf{\Sigma}$ from the uncertainty of each prediction $\sigma_{ij}$, calculated as the square root of the variance in~\eqref{eqn:gpr_var}. This yields a matrix $\boldsymbol{Z}$ alongside its corresponding trustworthiness indicator matrix $\mathbf{\Sigma}$. The detailed procedure is outlined in lines 13 through 26 of Algorithm~\ref{alg:mc_algo}.

\subsubsection{Constrained Nuclear Norm Minimization}
To capture the global smooth trend of the REM, we employ constrained nuclear norm minimization. This approach minimizes the nuclear norm as a proxy for the matrix rank, thereby enforcing the smooth spatial transitions typically inherent in large-scale propagation maps~\cite{sun2022propagation}. Let the SF observation matrix and its corresponding standard deviation matrix from the previous subsection be denoted as $\boldsymbol{Z}, \boldsymbol{\Sigma} \in \mathbb{R}^{N_\text{r} \times N_\text{c}}$, where $N_\text{r}$ and $N_\text{c}$ represent the number of rows and columns, respectively. The entries of these matrices are $\boldsymbol{Z}_{ij}$ and $\sigma_{ij}$. The refined smooth matrix $\hat{\boldsymbol{Z}}$ is subsequently recovered by solving the following constrained optimization problem~\cite{sun2022propagation}:
\begin{eqnarray}
\label{eqn:matrix_completion}
    \underset{\hat{\boldsymbol{Z}} \,\in\, \mathbb{R}^{N_\text{r} \times N_\text{c}}}{\operatorname{min}} & \|\hat{\boldsymbol{Z}}\|_* \nonumber \\
\text { subject to } & \abs{\hat{\boldsymbol{Z}}_{i j}-\boldsymbol{Z}_{i j}} \leq \alpha\sigma_{ij}, \quad \forall(i, j)
\end{eqnarray}
where $\|\cdot\|_*$ denotes the nuclear norm and $\alpha$ is a tuning parameter, detailed in Section~\ref{subsec:hyperparam}, that regularizes the deviation from the GPR estimates $\boldsymbol{Z}$. A large $\alpha$ promotes a lower-rank solution by penalizing the nuclear norm of $\hat{\boldsymbol{Z}}$, potentially filtering out localized spatial variations. Conversely, as $\alpha \to 0$, the optimization favors data fidelity, preserving the original GPR values. We solve~\eqref{eqn:matrix_completion} using an iterative approach, as detailed in Algorithm~\ref{alg:mc_func}, which employs the $\text{FindNuclearNormApprox}(\cdot)$ function~\cite{Shabat2023} to project the input matrix onto a target nuclear norm.
\begin{algorithm}[!t]
    \caption{Constrained nuclear norm minimization.}
    \label{alg:mc_func}
        \begin{algorithmic}[1]
            \Procedure{$\textnormal{NuclearNormMin}$}{}
            \State $\textbf{Input:}\,\, \textbf{Z} \!\in\! \mathbb{R}^{N_\text{r} \times N_\text{c}}, \boldsymbol{\Sigma} = (\sigma^2_{ij}) \in \mathbb{R}^{N_\text{r} \times N_\text{c}}, \alpha\in \mathbb{R}, T_{\lambda}\in \mathbb{R}$
            \State $\textbf{Initialize:}$
            \State $\hat{\textbf{Z}} \gets \textbf{Z},  \lambda \gets \|\textbf{Z}\|_*$
            \State $\lambda_{\text{min}} \gets 0, \lambda_{\text{max}} \gets \|\textbf{Z}\|_*, \delta_\lambda \gets \infty$
            \State $ {\Omega} = \{(i, j) \in \mathbb{N} \times \mathbb{N} \mid i \leq N_\text{r}, j \leq N_\text{c}\}$
            \State $\textbf{Loop:}$
            \While{$\exists (i, j) \in {\Omega} \mid \abs{\hat{\text{Z}}_{ij}-\text{Z}_{i j}} \geq \alpha\sigma_{ij}\, \textbf{or}\, \delta_\lambda > T_{\lambda}$} 
            \State $\lambda_{\text{old}} \gets \lambda$
            \State $\lambda \gets (\lambda_{\text{min}}+\lambda_{\text{max}})/2$
            \State $\hat{\textbf{Z}} \gets \text{FindNuclearNormApprox}(\hat{\textbf{Z}}, \lambda)$\Comment{\cite{Shabat2023}}
            \If {$\exists (i, j) \in {\Omega} \mid \abs{\hat{\text{Z}}_{ij}-\text{Z}_{i j}} \geq \alpha\sigma_{ij}$}
            $\lambda_{\text{min}} \gets \lambda$
            \Else
            $\,\,\lambda_{\text{max}} \gets \lambda$
            \EndIf
            \State $\delta_\lambda \gets \abs{\lambda-\lambda_{\text{old}}}$
            \EndWhile
            \State $\textbf{Output:}\,\hat{\textbf{Z}}$
            \EndProcedure
        \end{algorithmic}
\end{algorithm}
\subsubsection{Deep-Shadow Decomposition and Dilation}
Following the acquisition of the GPR-predicted matrix \(\boldsymbol{Z}\) and its smooth counterpart \(\hat{\boldsymbol{Z}}_\text{mc}\) (derived via constrained nuclear norm minimization), we identify regions where the absolute difference between the two exceeds a predefined threshold, $T_\text{v}$. These regions are categorized as deep-shadowed zones with high spatial variance. Subsequently, a grayscale dilation operation is applied to the isolated deep-shadow matrix to propagate these extrema to neighboring spatial locations. This dilated component is then recombined with the smooth portion to produce the final REM, effectively ensuring that deep shadowing is accounted for in the surrounding neighborhood. The procedure is outlined in lines 5 through 9 of Algorithm~\ref{alg:mc_algo}.

\subsubsection{Prediction}
To predict at an unsampled coordinate $(\psi_0, \zeta_0, h_0)$, defined by its latitude, longitude, and altitude, 2D interpolation is performed across latitude and longitude. This process utilizes the discrete matrix grid centers as reference points and is implemented via the MATLAB $\mathtt{interp2}(\cdot)$ function. This step corresponds to line 11 of Algorithm~\ref{alg:mc_algo}.

\section{Antenna Pattern Calibration (APC) and Enhanced Propagation Modeling}
\label{sec:calib_rad_prop}
This section presents the UAV-specific APC framework. Section~\ref{subsec:calib_method} details the calibration methodology, while Section~\ref{subsec:calib_res} provides some examples of empirically obtained radiation patterns and associated technical insights.

\subsection{Methodology of Calibration and Propagation Modeling}
\label{subsec:calib_method}
The radiation pattern of an antenna is modified when operating in close proximity to a UAV. This modification is deterministic; thus, the auxiliary experiment $\mathcal{E}'$ can be used to model the calibrated UAV-antenna pattern, $\text{G}_{\text{uav}}^{\text{(APC)}}(\phi_\mathrm{r}, \theta_\mathrm{r})$, where $\theta_\mathrm{r}$ and $\phi_\mathrm{r}$ denote the receive elevation and azimuth angles, respectively. This characterization rests on two key assumptions: (i) the GS antenna radiation pattern remains consistent with its isolated measurements, and (ii) dominant multipath components are generated by the UAV body rather than external reflectors.

The first assumption is justified by the typical deployment of GSs in isolated, elevated positions. The second assumption is increasingly valid at higher UAV altitudes, where reflections from remote terrestrial objects are spatially filtered, leaving the UAV chassis as the primary scattering source. Under these conditions, the calibrated antenna pattern is formulated as:
\begin{equation}
\text{G}_{\text{uav}}^{\text{(APC)}}(\phi_\mathrm{r}, \theta_\mathrm{r}) = \left( \frac{4 \pi}{l_{\text{wave}}} \right)^2 \frac{d_{\mathrm{3D}}^2 \cdot \mathrm{\hat{A}}_{\text{uav}}(\phi_\mathrm{r}, \theta_\mathrm{r}, d_{\mathrm{3D}})}{\mathrm{G}_{\text{gs}}(\phi_\mathrm{t}, \theta_\mathrm{t})},
\label{eq:effective_patt_calc1}
\end{equation}
where the intermediate term $\mathrm{\hat{A}}_{\text{uav}}(\phi_\mathrm{r}, \theta_\mathrm{r}, d_{\mathrm{3D}})$ is given by:
\begin{equation}
\mathrm{\hat{A}}_{\text{uav}}(\phi_\mathrm{r}, \theta_\mathrm{r}, d_{\mathrm{3D}}) = \sqrt{\frac{\mathrm{P}_{\mathrm{Rx}}(\phi_\mathrm{r}, \theta_\mathrm{r}, d_{\mathrm{3D}})}{\mathrm{P}_{\mathrm{Tx}}}}.
\label{eqn:A_uav_calc1}
\end{equation}
The full derivation of \eqref{eq:effective_patt_calc1} is provided in Appendix~\ref{sec:app1}. In practice, $\text{G}_{\text{uav}}^{\text{(APC)}}(\phi_\mathrm{r}, \theta_\mathrm{r})$ calculation requires sufficient measurement samples at each receive angle; when absent, the model can revert to the antenna's baseline isolated radiation pattern.

To enhance the radio propagation model, we first quantify the discrepancy between the calibrated radiation pattern obtained from \(\mathcal{E}'\), and the initial isolated pattern, as follows:
\begin{equation}
\Delta \text{G}_{\text{uav}}^{\text{(dB)}}(\phi_{\mathrm{r}}, \theta_{\mathrm{r}}) = \text{G}_{\text{uav}}^{(\text{APC})(\text{dB})}(\phi_{\mathrm{r}}, \theta_{\mathrm{r}}) - \text{G}_{\text{uav}}^{\text{(dB)}}(\phi_{\mathrm{r}}, \theta_{\mathrm{r}}).
\end{equation}
Next, we revise the received power calculation for the calibrated radiation pattern, as shown below:
\begin{equation}
\hat{r}_{\mathrm{trpl}}^{(\text{APC})(\text{dB})} = \mathrm{P}_{\mathrm{Tx}}^{(\text{dB})} - \mathrm{PL}_{\mathrm{trpl}}^{(\text{dB})} + \Delta \text{G}_{\text{uav}}^{\text{(dB)}}(\phi_{\mathrm{r}}, \theta_{\mathrm{r}}).
\label{eqn:received_power_tworay_calibrated}
\end{equation}
The above equation enhances the radio propagation model by incorporating a deterministic correction term for antenna gain.

\subsection{Calibrated Patterns: Insights from Experimental Data}
\label{subsec:calib_res}
\begin{figure}[t!]
    \centering
    \begin{subfigure}{0.24\textwidth}
        \centering
        \includegraphics[width=\linewidth]{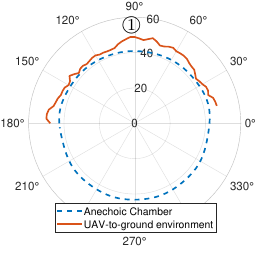}
        \caption{A3 (70~m Altitude)}
    \end{subfigure}
    \hfill
    \begin{subfigure}{0.24\textwidth}
        \centering
        \includegraphics[width=\linewidth]{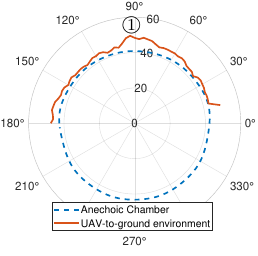}
        \caption{A5 (110~m Altitude)}
    \end{subfigure}
    \hfill
    \begin{subfigure}{0.24\textwidth}
        \centering
        \includegraphics[width=\linewidth]{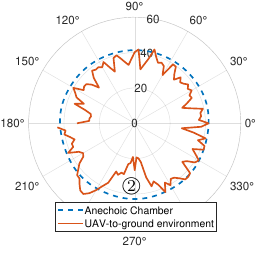}
        \caption{B2 (Fig.~\ref{fig:trajectories_afar}(b))}
    \end{subfigure}
    \hfill
    \begin{subfigure}{0.24\textwidth}
        \centering
        \includegraphics[width=\linewidth]{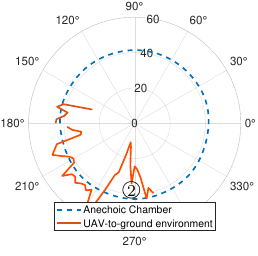}
        \caption{B3 (Fig.~\ref{fig:trajectories_afar}(c))}
    \end{subfigure}
    \hfill
    \begin{subfigure}{0.24\textwidth}
        \centering
        \includegraphics[width=\linewidth]{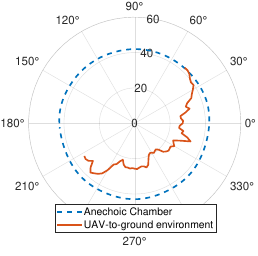}
        \caption{C1 (5 MHz Bandwidth)}
    \end{subfigure}
    \hfill
    \begin{subfigure}{0.24\textwidth}
        \centering
        \includegraphics[width=\linewidth]{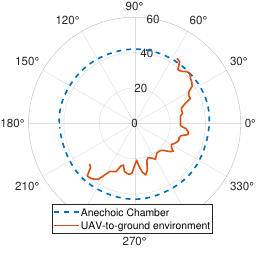}
        \caption{C4 (1.25 MHz Bandwidth)}
    \end{subfigure}
    \hfill
    \vspace{-3mm}
        \caption{Azimuth radiation patterns for six experiments from Table~\ref{tab:exp_list}. The solid line represents the radiation pattern in the experimental environment, while the dashed line represents the radiation pattern in the anechoic chamber. The subfigures are arranged for left-to-right comparison: (a) and (b) compare two \textit{heights}, (c) and (d) compare two \textit{trajectories}, and (e) and (f) compare two \textit{bandwidths}. All other conditions in each row remain identical, except for the one specified in the subcaptions. The plots are based on measurements taken only at low elevation angles (less than \(20^\circ\)).}
    \label{fig:ant_patt_azim}\vspace{-3mm}
\end{figure}

\begin{figure}[t!]
    \centering
    \begin{subfigure}{0.24\textwidth}
        \centering
        \includegraphics[width=\linewidth]{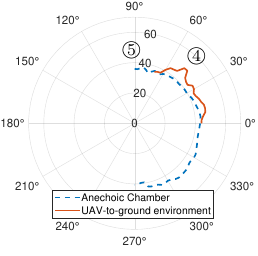}
        \caption{A3 (70~m Altitude)}
    \end{subfigure}
    \hfill
    \begin{subfigure}{0.24\textwidth}
        \centering
        \includegraphics[width=\linewidth]{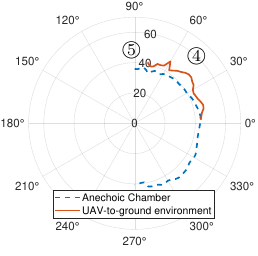}
        \caption{A5 (110~m Altitude)}
    \end{subfigure}
    \hfill
    \begin{subfigure}{0.24\textwidth}
        \centering
        \includegraphics[width=\linewidth]{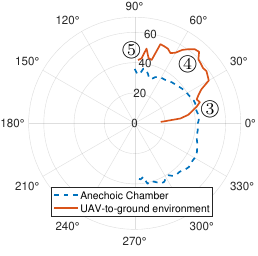}
        \caption{B2 (Fig.~\ref{fig:trajectories_afar}(b))}
    \end{subfigure}
    \hfill
    \begin{subfigure}{0.24\textwidth}
        \centering
        \includegraphics[width=\linewidth]{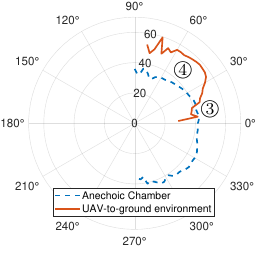}
        \caption{B3 (Fig.~\ref{fig:trajectories_afar}(c))}
    \end{subfigure}
    \hfill
    \begin{subfigure}{0.24\textwidth}
        \centering
        \includegraphics[width=\linewidth]{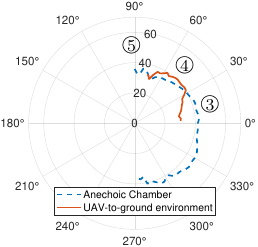}
        \caption{C1-C3 (5 MHz Bandwidth)}
    \end{subfigure}
    \hfill
    \begin{subfigure}{0.24\textwidth}
        \centering
        \includegraphics[width=\linewidth]{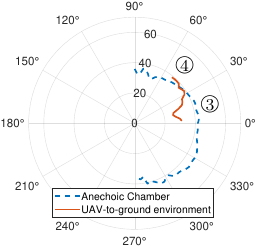}
        \caption{C4 (1.25 MHz Bandwidth)}
    \end{subfigure}
    \hfill
    \vspace{-3mm}
        \caption{Elevation radiation patterns for six experiments from Table~\ref{tab:exp_list}. The solid line represents the radiation pattern in the experimental environment, while the dashed line represents the radiation pattern in the anechoic chamber. The subfigures are arranged for left-to-right comparison: (a) and (b) compare two \textit{heights}, (c) and (d) compare two \textit{trajectories}, and (e) and (f) compare two \textit{bandwidths}. All other conditions of each row remain identical, except for the one specified in the subcaptions. The only exception is (e), which includes experiments from three different altitudes.}
    \label{fig:ant_patt_elev}\vspace{-3mm}
\end{figure}

Using the measurement data from each of the \( 15 \) experiments listed in Table~\ref{tab:exp_list} as training data \(\mathcal{E}'\), we compute the calibrated radiation pattern of the UAV-mounted antenna. We show representative results from two experiments of each dataset in Fig.~\ref{fig:ant_patt_azim} and Fig.~\ref{fig:ant_patt_elev}.

The calibrated azimuth radiation patterns shown in Fig.~\ref{fig:ant_patt_azim} provide valuable insights. First, although the antennas in the anechoic chamber appear omnidirectional, this is not the case in field experiments. Second, each data collection environment, such as the one depicted in Fig.~\ref{fig:uav_trajectory} (LTE I/Q dataset), results in distinct radiation patterns in Fig.~\ref{fig:ant_patt_azim}(a) and Fig.~\ref{fig:ant_patt_azim}(b). For example, the antenna gain increases at around \(90^\circ\) in both Fig.~\ref{fig:ant_patt_azim}(a) and Fig.~\ref{fig:ant_patt_azim}(b), as indicated by \raisebox{.5pt}{\textcircled{\raisebox{-.9pt}{1}}}. In the AFAR environment (Fig.~\ref{fig:trajectories_afar}), the antenna pattern shows a blockage around \(270^\circ\) in Fig.~\ref{fig:ant_patt_azim}(c) and Fig.~\ref{fig:ant_patt_azim}(d), as highlighted by \raisebox{.5pt}{\textcircled{\raisebox{-.9pt}{2}}}. Similarly, a unique radiation pattern is observed in the multi-band dataset, as shown in Fig.~\ref{fig:ant_patt_azim}(e) and Fig.~\ref{fig:ant_patt_azim}(f). While Fig.~\ref{fig:ant_patt_azim}(a) and (b) vary altitude, (c) and (d) vary trajectory, and (e) and (f) vary bandwidth, the side-by-side comparison of the azimuth patterns confirms that the effective radiation pattern remains largely unaffected by altitude, bandwidth, or trajectory. However, it does vary with the experimental setup, including the characteristics of the GS, the surrounding environment, and the UAV used for sensing. These factors differ across datasets, leading to variations in the radiation patterns across different rows in Fig.~\ref{fig:ant_patt_azim}.

Fig.~\ref{fig:ant_patt_elev} illustrates the elevation radiation patterns, which, like the azimuth radiation patterns, show similar dependence on the experimental environment, altitude, trajectory, and bandwidth. However, a consistent trend is observed in Fig.~\ref{fig:ant_patt_elev}(a) to Fig.~\ref{fig:ant_patt_elev}(f) across different elevation angles. At lower elevation angles, the radiation pattern based on field measurements shows a significant decrease compared to the anechoic chamber measurements in Fig.~\ref{fig:ant_patt_elev}(c) to Fig.~\ref{fig:ant_patt_elev}(f), as marked by \raisebox{.5pt}{\textcircled{\raisebox{-.9pt}{3}}}. As the elevation angle increases, the antenna gain surpasses that of the anechoic chamber, as highlighted by \raisebox{.5pt}{\textcircled{\raisebox{-.9pt}{4}}}. At \(90^\circ\), when the UAV is directly above the ground source, the pattern returns to the anechoic chamber gain, as indicated by \raisebox{.5pt}{\textcircled{\raisebox{-.9pt}{5}}}.

The increased antenna gain in Fig.~\ref{fig:ant_patt_elev}(c) and Fig.~\ref{fig:ant_patt_elev}(d), at medium elevation angles, highlighted by \raisebox{.5pt}{\textcircled{\raisebox{-.9pt}{4}}}, is attributed to strong reflections from surfaces near the GS antenna. \textit{The likelihood of reflection is greater in this experimental scenario} because the UGV antenna was positioned at a relatively low altitude, which enhances the chances of reflections from the UGV surface. In contrast, the BS antenna is expected to interact less with any nearby surfaces. The reduction in gain at lower altitudes, highlighted by \raisebox{.5pt}{\textcircled{\raisebox{-.9pt}{3}}}, is likely due to \textit{increased susceptibility to LoS blocking and/or destructive interference from reflected components} at lower elevation angles. To validate these interpretations, we conducted MATLAB-based ray-tracing simulations, as detailed in Appendix~\ref{sec:app2}. By incorporating a 3D UAV object and GS, we reproduced the multipath environments responsible for the observed constructive and destructive interference patterns.

\section{Experimental Results and Analysis}
\label{sec:num_results}
In this section, we present the experimental results for 3D REM reconstruction. The experimental setup is provided in Section~\ref{subsec:hyperparam}. The parametric sensitivity of REM reconstruction accuracy across elevation angle, UAV altitude, and spectral bandwidth is evaluated in Sections~\ref{sec:ana_elevation} through \ref{sec:ana_ban}. Finally, a comparative analysis of REM reconstruction methods, incorporating Kriging variants, APC framework, and our proposed MC-assisted GPR, is provided in Sections~\ref{sec:ana_kriging} through \ref{sec:ana_gpr}.

\subsection{Experimental Setup and Evaluation Criteria}
\label{subsec:hyperparam}
\subsubsection{Train-Test Split}
As established in Section~\ref{sec:pro_set}, the test experiment $\mathcal{E}$ and training experiment $\mathcal{E}'$ are from separate trials. From the target experiment $\mathcal{E}$, only a limited number of samples $\mathrm{M}$ are available for REM reconstruction. To facilitate effective calibration, a distinct training experiment $\mathcal{E}'$ is selected from Table~\ref{tab:exp_list} based on its proximity to the parameters of $\mathcal{E}$. Specifically, $\mathcal{E}'$ is chosen as the adjacent experiment with the nearest vertical displacement (e.g., a 20~m altitude separation) or the closest spectral configuration (e.g., a 1.25~MHz bandwidth offset), depending on data availability.
\subsubsection{Hyperparameter Settings}
\begin{figure}
    \centering
        \includegraphics[width=0.7\linewidth]{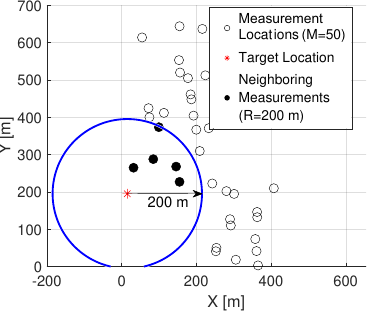}
    \caption{Illustration of the number of available sparse samples \(\mathrm{M}\) and sample selection radius \(\mathrm{R}\) used in Kriging. For predicting at the center of the circle, Kriging utilizes only the samples within the circle.}
    \label{fig:fixed_radius_and_points}
    \vspace{-4mm}
\end{figure}

Given $\mathrm{M}$ sparse samples, Kriging interpolation utilizes a selection radius $\mathrm{R}$ centered at the prediction location, as illustrated in Fig.~\ref{fig:fixed_radius_and_points}. This localized approach limits the number of active samples, mitigating the potential performance degradation typically associated with high-dimensional Kriging matrices. In this study, $\mathrm{R}$ was set to either 70~m or 200~m, where $\mathrm{R}=200$~m was found to fully exploit the spatial interpolation potential of the Kriging framework. The hyperparameters for Algorithm~\ref{alg:mc_algo} and Algorithm~\ref{alg:mc_func} were configured as $\alpha = 1$, $T_\text{v} = 1$, and $T_\lambda = 20$.
\subsubsection{Evaluation Criteria}
For a given set of $\mathrm{M}$ samples, the REM reconstruction accuracy is evaluated using the Root Mean Square Error (RMSE). To ensure statistical significance, the evaluation process was repeated for 5,000 Monte Carlo iterations. In each iteration, $\mathrm{M}$ measurement locations were randomly sampled from the test experiment $\mathcal{E}$. The median RMSE across these iterations is reported as the primary metric for performance comparison.

\subsection{Sensitivity of REM Accuracy to UAV Elevation Angle}
\label{sec:ana_elevation}
\begin{figure}[!ht]
\centerline{
\includegraphics[width=\linewidth]{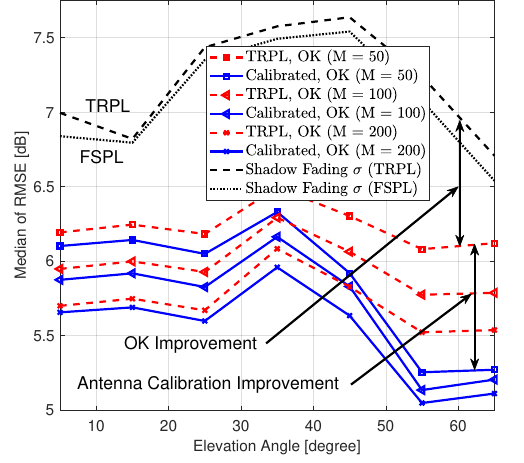}
}
\caption{REM reconstruction accuracy, measured by the median RMSE across elevation angles, comparing the (TRPL, OK) and (APC, TRPL, OK) models. The SF standard deviation $\sigma$, estimated via TRPL and FSPL, is also illustrated. The reduction in RMSE when using (TRPL, OK) relative to the TRPL-estimated $\sigma$ is identified as the ``OK Improvement", while the further performance gain is attributed to APC.}
\label{fig:altitudes_sung_5_reason}
\vspace{-3mm}
\end{figure}

Fig.~\ref{fig:altitudes_sung_5_reason} presents the REM reconstruction performance, in terms of the median RMSE, across elevation angles ranging from \(5^\circ\) to \(65^\circ\) using the LTE I/Q dataset.
The reconstruction performance is measured for TRPL with OK, both with and without APC (solid blue and dashed red lines, respectively). These results are presented for $\mathrm{M} \in \{50, 100, 200\}$ and $\mathrm{R}=200$~m. The SF standard deviation $\sigma$, estimated from both TRPL and FSPL, is also plotted as a baseline.

The results indicate that the TRPL-estimated $\sigma$ varies significantly with elevation: it initially decreases between $5^\circ$ and $15^\circ$, reaches a peak at $45^\circ$, and subsequently declines rapidly at steeper elevation angles. Notably, the reconstruction performance of OK (indicated by red dashed lines) does not always align with TRPL-estimated $\sigma$. Specifically, at higher elevation angles such as $65^\circ$, the performance gain from OK (labeled as ``OK improvement") is comparatively lower.

\begin{figure}[t!]
\centerline{
\includegraphics[width=\linewidth]{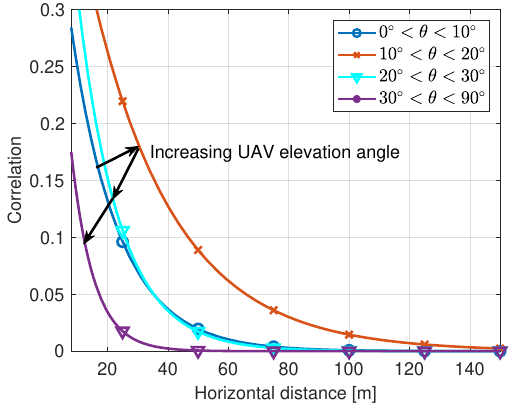}
}
\caption{Shadow fading correlation profile as a function of horizontal distance for varying UAV elevation angles. Correlation strength initially increases with elevation angle, then declines and eventually reaches near-zero correlation.} % Results are derived from the experimental LTE I/Q dataset.
\label{fig:elevation_group_corr}
\vspace{-4mm}
\end{figure}

To investigate the efficacy of OK across elevations, the spatial correlation versus horizontal distance for varying elevation angles is plotted in Fig.~\ref{fig:elevation_group_corr}.
The correlation exhibits non-monotonic behavior, increasing up to $20^\circ$ before declining toward negligible levels at steeper angles. This explains the marginal OK gains observed in Fig.~\ref{fig:altitudes_sung_5_reason} at $65^\circ$. It also aligns with the observed trend of lower OK improvement at small elevation angles ($<10^\circ$), followed by increased gains as the elevation angle increases toward the mid-range.

The characteristic profile of the TRPL $\sigma$ with respect to elevation can be interpreted through the physical propagation environment. At lower elevation angles, the UAV is more susceptible to LoS blockage and a higher density of terrestrial multipath reflections, resulting in higher SF variance at $5^\circ$ compared to $15^\circ$. As the elevation angle increases, the UAV clears most terrestrial obstacles, leading to a more LoS-dominant environment and a subsequent reduction in SF variance. The SF variance peak centered at $45^\circ$ is attributed to the GS's dipole antenna pattern, which introduces larger received power fluctuations in this region. As the elevation increases further, the GS's influence diminishes due to the dipole's radiation nulls, leading to more consistent SF behavior in terms of variance.

\subsection{Sensitivity of REM Accuracy to UAV Altitude}
\label{subsec:ana_alt}
\begin{figure}[!t]
    \centering
    \begin{subfigure}{0.48\textwidth}
        \centering
        \includegraphics[width=\linewidth]{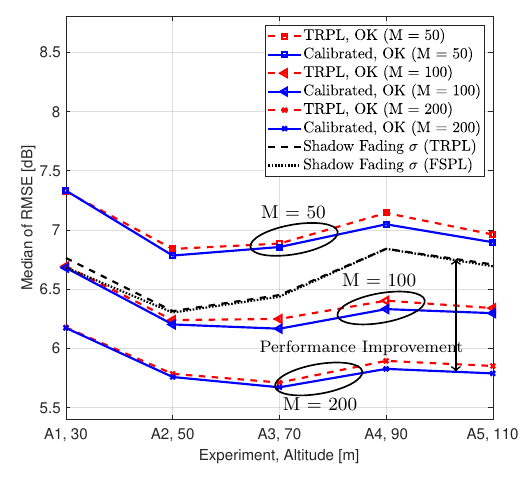}
        \caption{\(\mathrm{R}=70\)~m}
    \end{subfigure}
    %\hfill
    \begin{subfigure}{0.48\textwidth}
        \centering
        \includegraphics[width=\linewidth]{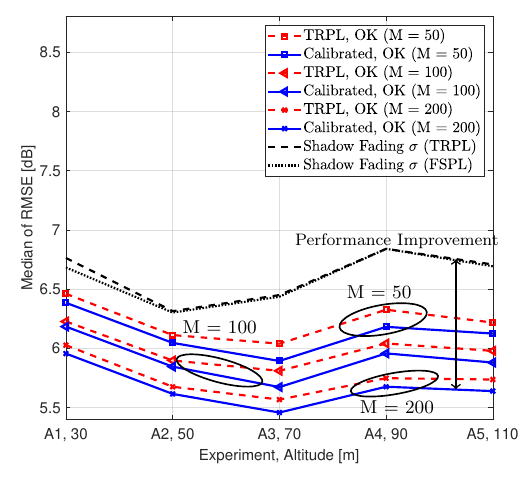}
        \caption{\(\mathrm{R}=200\)~m}
    \end{subfigure}
    \caption{Performance of RSRP prediction across five experiments with increasing altitudes from the LTE I/Q dataset. The results are presented for TRPL with OK, both with and without APC (solid blue and dashed red lines, respectively), computed for (a) \( \mathrm{R} = 70 \, \text{m} \) and (b) \( \mathrm{R} = 200 \, \text{m}.\) RMSE performance improvement compared to shadow fading \(\sigma\) is marked.} % as overall ``performance improvement".This gain increases with altitude.
    \label{fig:altitudes_sung_5}
    \vspace{-6mm}
\end{figure}

Fig.~\ref{fig:altitudes_sung_5} illustrates the median RMSE results for UAV altitudes ranging from 30~m to 110~m, derived from the experimental LTE I/Q dataset.
Performance is evaluated for TRPL with OK, both with and without APC. The estimated SF standard deviation $\sigma$ is also plotted as a baseline. 
We observe that $\sigma$ exhibits a distinct tri-phasic behavior with respect to altitude: an initial decrease, followed by an increase that peaks at 70~m, and a subsequent decline at higher altitudes. This trend is consistent with the behavior of \(\sigma\) with elevation angle in Fig.~\ref{fig:altitudes_sung_5_reason}. The physical interpretation provided in Section~\ref{sec:ana_elevation}~remains applicable here, as altitude and elevation angle are intrinsically linked; the initial decline results from increased LoS dominance, the peak at 90~m is attributed to the GS's dipole radiation pattern, and the final decline occurs as the UAV moves into the dipole's upper nulls.

\begin{figure}[!t]
\centerline{
\includegraphics[width=\linewidth]{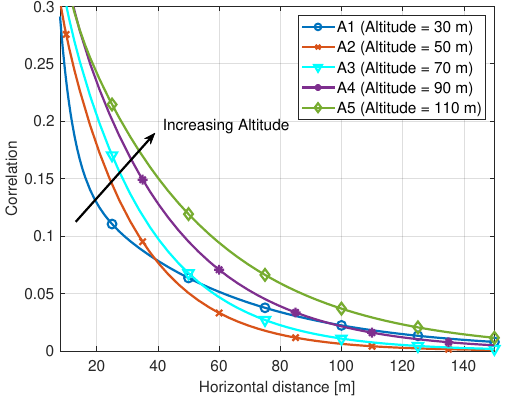}
}
\caption{Correlation profile across horizontal distance for five experiments at increasing altitudes from the LTE I/Q dataset. At higher altitudes, the correlation decays at a slower rate.}
\label{fig:corr_compare_heights}
\vspace{-4mm}
\end{figure}

The reconstructed REM performance is represented by the dashed red lines and the solid blue lines (with APC). Both models reduce the estimation error below the raw SF $\sigma$, with the delta identified as the ``performance improvement" provided by the OK algorithm. As shown in Fig.~\ref{fig:altitudes_sung_5}(b), this improvement increases with altitude. To investigate this altitude-dependent gain, the spatial correlation profiles for discrete altitudes are plotted in Fig.~\ref{fig:corr_compare_heights}. 
As altitude increases, the spatial correlation decays more slowly, indicating that correlation-based methods like OK are more effective at higher altitudes. This behavior aligns with the elevation-based correlation trends in Fig.~\ref{fig:elevation_group_corr}, as increasing altitude shifts the average elevation angle of the trajectory from the low-elevation regime ($0^\circ < \theta < 10^\circ$) toward the mid-elevation regime ($10^\circ < \theta < 20^\circ$), where spatial correlation is stronger.

Consistent with the tri-phasic behavior of $\sigma$, the reconstructed REM performance exhibits a similar tri-phasic trend across altitudes. This behavior is not isolated to this specific dataset; a study by Ivanov~\textit{et al.}~\cite{ivanov2023limited} also demonstrates a similar pattern in their results. Furthermore, we observe this consistent tri-phasic characteristic in results derived from other experimental datasets, as detailed in Appendix~\ref{sec:app3}.

Furthermore, Fig.~\ref{fig:altitudes_sung_5}(a) shows that with a limited sample size ($\mathrm{M} = 50$) and a small selection radius ($\mathrm{R} = 70$~m), OK can actually perform worse than simple TRPL-based estimation. However, increasing the selection radius to $\mathrm{R} = 200$~m, as shown in Fig.~\ref{fig:altitudes_sung_5}(b), significantly enhances the performance, confirming that a larger neighborhood is required for OK for optimal performance under limited sampling, \(\mathrm{M} \leq 200\).

\subsection{Sensitivity of REM Accuracy to Channel Bandwidth}
\label{sec:ana_ban}
\begin{figure}[t!]
\centerline{
      \includegraphics[width=\linewidth]{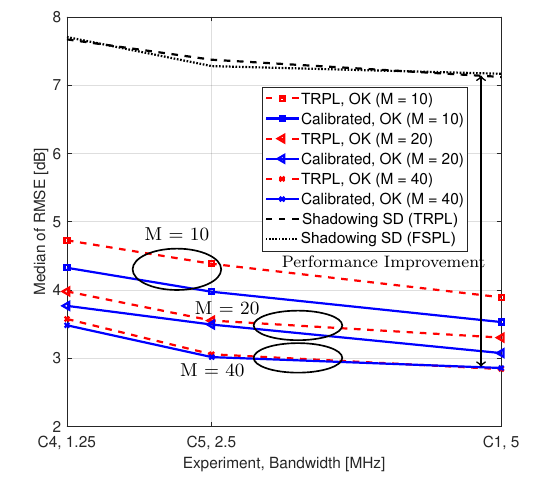}
}
\caption{RSRP prediction error across three experiments with increasing bandwidths from the multi-band dataset.}
\label{fig:altitudes_cole_bandwidths}
\vspace{-4mm}
\end{figure}

Fig.~\ref{fig:altitudes_cole_bandwidths} presents the median RMSE results for three experiments with bandwidths of \( 1.25 \, \text{MHz} \), \( 2.5 \, \text{MHz} \), and \( 5 \, \text{MHz} \) from the multi-band dataset.
The altitudes for these experiments were fixed at 40~m, and the UAV trajectories were identical. The figure plots the performance for TRPL with OK, both with and without APC (solid blue and dashed red lines, respectively). The observed trend indicates that the prediction error decreases as the bandwidth increases. For instance, the median RMSE is approximately 0.8~dB lower when using TRPL with OK at a 5~MHz bandwidth compared to a 1.25~MHz bandwidth. This behavior is expected, as narrower bandwidths are more susceptible to frequency-selective fading, which increases SF variance. For this evaluation, the OK configuration was maintained at $\mathrm{R} = 200$~m.

\subsection{Comparative REM Accuracy across Kriging Variants}
\label{sec:ana_kriging}
\begin{figure}[t!]
    \centering
    %\hfill
    \begin{subfigure}{0.48\textwidth}
        \centering
        \includegraphics[width=\linewidth]{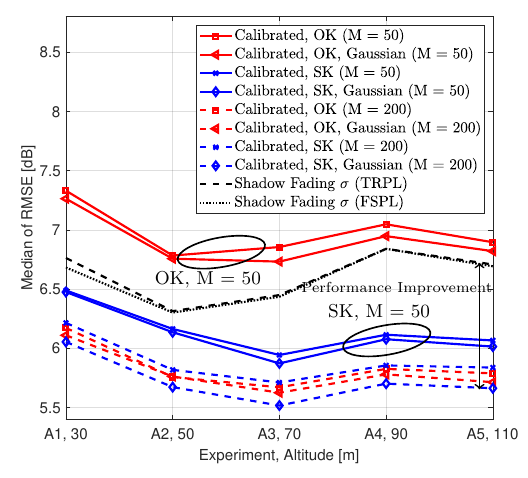}
        \caption{\(\mathrm{R}=70\)~m}
    \end{subfigure}
    \begin{subfigure}{0.48\textwidth}
        \centering
        \includegraphics[width=\linewidth]{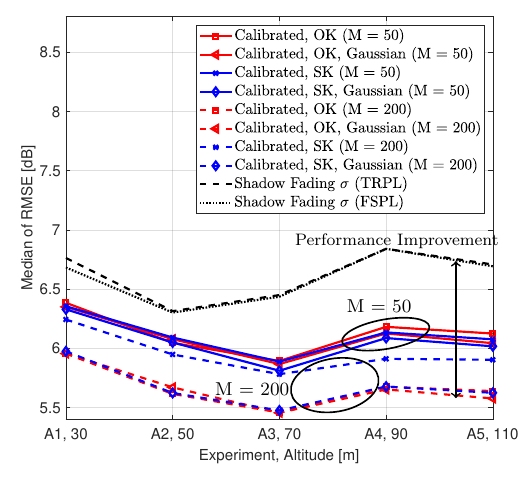}
        \caption{\(\mathrm{R}=200\)~m}
    \end{subfigure}
    \caption{Performance comparison of OK, SK, and their trans-Gaussian variants across five experiments in LTE I/Q dataset, using APC, for (a) \( \mathrm{R} = 70 \, \text{m} \) and (b) \( \mathrm{R} = 200 \, \text{m}.\) Trans-Gaussian SK shows the best overall performance.}
    \label{fig:altitudes_sung_5_variations}
    \vspace{-4mm}
\end{figure}

Fig.~\ref{fig:altitudes_sung_5_variations} illustrates the comparative performance of OK, SK, and their trans-Gaussian variants.
The results, shown in Fig.~\ref{fig:altitudes_sung_5_variations}(a) and Fig.~\ref{fig:altitudes_sung_5_variations}(b), represent the REM reconstruction result using TRPL with a Kriging variant for \( \mathrm{R} = 70 \, \text{m} \) and \( \mathrm{R} = 200 \, \text{m} \), respectively. In Fig.~\ref{fig:altitudes_sung_5_variations}(a), for \( \mathrm{R} = 70 \, \text{m} \) and \( \mathrm{M} = 50 \), SK provides a performance boost of approximately \( 1 \, \text{dB} \) over OK, demonstrating the effectiveness of SK when a small number of measurements are available. At \( \mathrm{M} = 200 \), the performance gain of SK diminishes, while trans-Gaussian SK provides the best performance across all altitudes. In Fig.~\ref{fig:altitudes_sung_5_variations}(b), SK performs relatively poorly for \( \mathrm{R} = 200 \, \text{m} \) and \( \mathrm{M} = 200 \), indicating that as the number of available samples increases, OK becomes a better option. However, trans-Gaussian SK maintains performance at a level comparable to OK. On the other hand, the performance gain from employing trans-Gaussian OK is negligible, with results nearly identical to standard OK.

\subsection{REM Accuracy Enhancement via Antenna Calibration}
\label{sec:ana_antenna}
The impact of APC on REM reconstruction accuracy was introduced in previous subsections, specifically within the contexts of Fig.~\ref{fig:altitudes_sung_5} and Fig.~\ref{fig:altitudes_cole_bandwidths}. In Fig.~\ref{fig:altitudes_sung_5}, the training experiment $\mathcal{E}'$ was characterized by a 20~m vertical displacement from the target experiment $\mathcal{E}$. Conversely, in Fig.~\ref{fig:altitudes_cole_bandwidths}, the separation was defined in the frequency domain, with bandwidth offsets of 1.25~MHz and 2.5~MHz.
Fig.~\ref{fig:altitudes_sung_5_reason} demonstrates that the most significant performance variation between the baseline TRPL and APC occurs at higher elevation angles. This confirms that the error reduction results from the calibrated radiation pattern, which becomes increasingly distinguishable at steeper elevation angles, as illustrated by the calibrated patterns in Fig.~\ref{fig:ant_patt_elev}(a) and Fig.~\ref{fig:ant_patt_elev}(b).

Extending the analysis of altitudinal and spectral separation, we observe that the antenna pattern can also be effectively learned from an experiment $\mathcal{E}'$ employing a distinct horizontal trajectory. The results of this cross-trajectory antenna calibration are provided in Appendix~\ref{sec:app4}. Finally, as discussed in Appendix~\ref{sec:app4}, in scenarios where $\mathcal{E}'$ provides insufficient sampling density for a specific target direction, the framework should revert to the baseline antenna pattern.

\subsection{REM Accuracy Enhancement via MC-Assisted GPR for Deep-shadow Extraction}
\label{sec:ana_gpr}
\begin{figure}[t!]
    \centering
    \begin{subfigure}{0.23\textwidth}
        \centering
        \includegraphics[width=\linewidth]{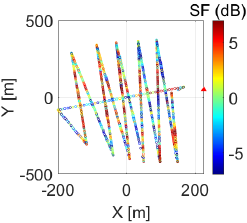}
        \caption{Ground-truth}
    \end{subfigure}
    \begin{subfigure}{0.23\textwidth}
        \centering
        \includegraphics[width=\linewidth]{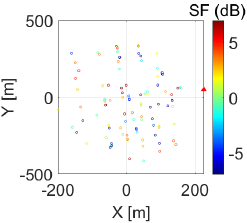}
        \caption{Samples (M=100)}
    \end{subfigure}
    \begin{subfigure}{0.23\textwidth}
        \centering
        \includegraphics[width=\linewidth]{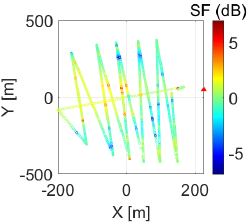}
        \caption{GPR}
    \end{subfigure}
    \begin{subfigure}{0.23\textwidth}
        \centering
        \includegraphics[width=\linewidth]{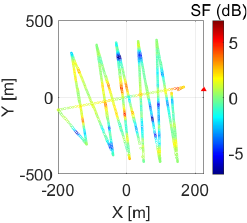}
        \caption{MC-assisted GPR}
    \end{subfigure}
    \vspace{-1mm}
        \caption{SF map reconstruction results: (a) ground-truth SF map, (b) observed sparse samples (M=100), (c) standard GPR prediction, and (d) MC-assisted GPR prediction. The map in (d) utilizes grayscale dilation to effectively spread deep-shadow regions.}
    \label{fig:sf_map_mc}\vspace{-4mm}
\end{figure}

In this subsection, we evaluate the performance of the proposed MC-Assisted GPR, which is designed to enhance deep-shadowed regions. Using the LTE I/Q experimental dataset, representative SF maps generated by the proposed algorithm and standard GPR are compared in Fig.~\ref{fig:sf_map_mc}.
As illustrated, the standard GPR prediction in Fig.~\ref{fig:sf_map_mc}(c) tends to treat deep-shadow observations (indicated by dark blue or red regions) as isolated singular points. Standard Kriging and GPR algorithms inherently attempt to ``erase'' or smooth these extrema to minimize global variance, preventing neighboring locations from benefiting from these deep-shadow observations. This oversmoothing is counterintuitive in environments where this effect is spatially continuous. In contrast, the proposed MC-assisted GPR specifically identifies these deep-shadowed points and propagates their influence through the local neighborhood using grayscale dilation, as shown in Fig.~\ref{fig:sf_map_mc}(d).

\begin{figure}[!t]
\centerline{
\includegraphics[width=\linewidth]{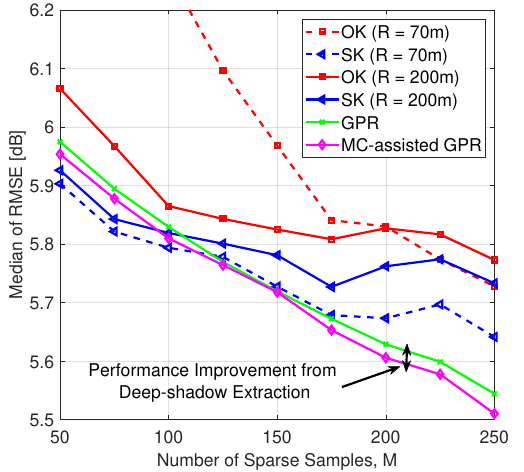}
}
\caption{Comparison of REM reconstruction performance among OK, SK, GPR, and the proposed MC-assisted GPR. In all cases, the standard TRPL serves as the baseline radio propagation model, with the respective algorithms used to reconstruct the SF residuals.}
\label{fig:mc_performance}
\vspace{-4mm}
\end{figure}

Fig.~\ref{fig:mc_performance} presents the REM reconstruction error, measured in median RMSE, for the proposed algorithm compared to OK, SK, and baseline GPR.
All algorithms reconstruct the SF maps using TRPL as the large-scale path loss model. The results indicate that GPR generally outperforms OK and SK, as the latter methods require the optimization of the selection radius $\mathrm{R}$ for a given sampling density $\mathrm{M}$. At very low sampling densities, SK provides the most accurate results; however, it lacks scalability as $\mathrm{M}$ increases, allowing OK to eventually match its performance. GPR, on the other hand, provides reliable and consistent performance gains with increasing $\mathrm{M}$. Because GPR explicitly accounts for measurement noise, it maintains stability even with a high number of observations, unlike SK, which may suffer from matrix ill-conditioning.

Finally, the proposed MC-assisted method consistently enhances GPR performance by effectively characterizing the deep-shadow zones. However, this improvement increases computational complexity. Unlike standard GPR, which can perform localized inference at arbitrary points without generating a global map, the MC-assisted approach requires a nuclear norm minimization operation over the full spatial matrix. Furthermore, while GPR operates on continuous spatial coordinates, the matrix-based MC framework necessitates location quantization, meaning the resolution and computational cost are intrinsically linked to the grid density.

\section{Conclusion}
\label{sec:conclusion}
This paper presents a comprehensive characterization of 3D REM reconstruction using real-world UAV measurements, providing fundamental insights into the physical and platform-specific drivers of estimation accuracy. We have demonstrated that the impact of UAV altitude on reconstruction fidelity follows a distinct tri-phasic trend, whereas increased spectral bandwidth consistently enhances reconstruction performance. Furthermore, our analysis reveals that shadowing variance is intrinsically linked to elevation angles; lower elevation angles yield higher variance due to reduced LoS dominance, while steep elevation angles experience lower shadowing variance as the UAV transitions into the antenna's dipole null regions. From a practical perspective, our findings provide a predictive framework for identifying spatial regions where signal strength fluctuates rapidly. This enables denser measurement collection in these regions for efficient spectrum sensing, avoiding exclusive reliance on computationally expensive deep learning-based trajectory planning.

Electromagnetic distortions induced by the UAV airframe cause consistent variations in the radiation pattern that are empirically modeled through our APC framework. By utilizing this calibrated pattern, REM reconstruction accuracy is significantly improved. Additionally, we have demonstrated that sparse samples in rapidly varying shadowing regions are often filtered or ``erased" by standard reconstruction algorithms. By employing a two-layer shadowing extraction framework, MC-assisted GPR, we successfully detected these regions and restored the sudden spatial transitions. Ultimately, these insights for single-source scenarios serve as essential building blocks for modeling more complex, multi-source 3D radio environments in future cognitive radio networks.

\section*{Acknowledgments}
We use the publicly available field measurement data and the preprocessing scripts~\cite{IEEEDataPort,IEEEDataPort_2,IEEEDataPort_3,IEEEDataPort_4} published by the NSF AERPAW~\cite{9061157,gurses2025digital} platform. This research is supported in part by the NSF awards CNS-1939334 and CNS-2450593, and the INL Laboratory Directed Research Development (LDRD) Program under BMC No. 264247, Release No. 26 on BEA’s Prime Contract No. DE-AC07-05ID14517.
\renewcommand{\thefootnote}{}
\footnote{\small Mushfiqur Rahman, \.{I}smail G\"{u}ven\c{c}, Chau-Wai Wong, and Mihail Sichitiu are with the Department of Electrical and Computer Engineering, North Carolina State University, Raleigh, NC 27606, USA (e-mail: mrahman7@ncsu.edu; iguvenc@ncsu.edu; chauwai.wong@ncsu.edu; mlsichit@ncsu.edu).}
\footnote{\small Sung Joon Maeng was previously with the Department of Electrical and Computer Engineering, North Carolina State University. He is now with the Department of Electrical and Electronic Engineering, Hanyang University, Ansan 15588, South Korea (e-mail: sjmaeng@hanyang.ac.kr).}
\footnote{\small Jason A. Abrahamson and Arupjyoti Bhuyan are with the Idaho National Laboratory, Idaho Falls, ID 83402, USA (e-mail: jason.abrahamson@inl.gov; arupjyoti.bhuyan@inl.gov).}

\appendices
\section{Analytical Derivation of the UAV-specific APC Equation}
\label{sec:app1}
The received power at a UAV-mounted antenna is the superposition of $S$ discrete multipath components. In the standard TRPL model, $S=2$, comprising the LoS and the ground-reflected ray. However, for an antenna integrated onto a UAV chassis, $S$ is typically larger due to near-field reflections and diffractions from the UAV’s structural components. The total received power is expressed as:
\begin{equation}
\mathrm{P}_{\mathrm{Rx}}(\phi_\mathrm{r}, \theta_\mathrm{r}, d_{\mathrm{3D}}) = \mathrm{P}_{\mathrm{Tx}} \left| \sum_{i=1}^{S} \mathrm{A}_i \exp(j \tau_i) \right|^2,
\label{eq:total_power_effective}
\end{equation}
where \( \mathrm{P}_{\mathrm{Tx}} \) and \( \mathrm{P}_{\mathrm{Rx}} \) represent the transmitted and received powers in watts, respectively; $(\phi_\mathrm{r}, \theta_\mathrm{r}, d_{\mathrm{3D}})$ denote the receive azimuth, elevation, and 3D Euclidean distance relative to the GS, as illustrated in Fig.~\ref{fig:diagram_two-ray-model}. Here, $\mathrm{A}_i$ denotes the amplitude attenuation and $\tau_i$ represents the phase shift of the $i$-th ray.

The effective radiation pattern of the UAV-mounted antenna is a composite of the direct LoS ray ($i=1$) and all secondary rays that interact exclusively with the UAV body along the propagation path. To isolate these platform-specific effects from terrestrial multipath, we exclude the ground-reflected ray ($i=2$) and define a resultant ``UAV-effective" phasor, $\mathrm{A}_{\text{uav}} \exp(j \tau_{\text{uav}})$, as follows:
\begin{equation}
\mathrm{A}_{\text{uav}} \exp(j \tau_{\text{uav}}) = \sum_{\substack{i=1 \\ i \neq 2}}^{S} \mathrm{A}_i \exp(j \tau_i).
\end{equation}
This resultant phasor accounts for all scattering and diffraction contributions intrinsic to the UAV platform. Consequently, \eqref{eq:total_power_effective} can be simplified as follows:
\begin{equation}
\mathrm{P}_{\mathrm{Rx}}(\phi_\mathrm{r}, \theta_\mathrm{r}, d_{\mathrm{3D}}) = \mathrm{P}_{\mathrm{Tx}} \left| \sum_{i \in \{\text{uav}, 2\}} \mathrm{A}_i \exp(j \tau_i) \right|^2.
\label{eq:total_power_effective2}
\end{equation}

To obtain the radiation pattern of the UAV-mounted antenna, $\mathrm{A}_{\text{uav}}$, a practical estimation method is required. While anechoic chamber measurements provide high precision, they are cost-prohibitive and must be repeated for every structural customization of the UAV body. In-field estimation via~\eqref{eq:total_power_effective2} serves as a viable alternative. However, solving for $\mathrm{A}_{\text{uav}}$ remains challenging because $\mathrm{P}_{\mathrm{Rx}}$ represents power (magnitude squared), whereas the constituent components are complex phasors. A full reconstruction would typically require channel-sounding waveforms to resolve individual phase contributions.
Furthermore, at a carrier frequency of $3.5$~GHz, the wavelength ($\lambda \approx 8.6$~cm) makes the phase of the reflected ray highly sensitive to small variations in the environment. Given that natural grasslands are not perfectly planar and that slight inaccuracies in 3D distance measurements are inevitable, the deterministic calculation of the reflected ray's phase is often unreliable. However, since the reflectivity of vegetation is significantly lower than that of the metallic components typical of UAV chassis, the impact of the ground-reflected ray can be considered secondary in high-altitude scenarios. By neglecting the ground-reflected component, we can approximate the UAV-effective term from~\eqref{eq:total_power_effective2} as:
\begin{equation}
\mathrm{\hat{A}}_{\text{uav}}(\phi_\mathrm{r}, \theta_\mathrm{r}, d_{\mathrm{3D}}) \approx \sqrt{\frac{\mathrm{P}_{\mathrm{Rx}}(\phi_\mathrm{r}, \theta_\mathrm{r}, d_{\mathrm{3D}})}{\mathrm{P}_{\mathrm{Tx}}}}.
\label{eqn:A_uav_calc}
\end{equation}

The resulting $\hat{\mathrm{A}}_{\text{uav}}$ in \eqref{eqn:A_uav_calc} represents the total path gain, which inherently contains the spatial characteristics of the UAV-mounted antenna. To isolate the UAV-antenna radiation pattern, $\text{G}_{\text{uav}}^{\text{(APC)}}(\phi_\mathrm{r}, \theta_\mathrm{r})$, we normalize this quantity by accounting for the 3D free-space path loss and the GS antenna gain:
\begin{equation}
\text{G}_{\text{uav}}^{\text{(APC)}}(\phi_\mathrm{r}, \theta_\mathrm{r}) = \left( \frac{4 \pi}{l_{\text{wave}}} \right)^2 \frac{d_{\mathrm{3D}}^2 \cdot \mathrm{\hat{A}}_{\text{uav}}(\phi_\mathrm{r}, \theta_\mathrm{r}, d_{\mathrm{3D}})}{\mathrm{G}_{\text{gs}}(\phi_\mathrm{t}, \theta_\mathrm{t})},
\label{eq:effective_patt_calc}
\end{equation}
where $l_{\text{wave}}$ is the wavelength, $\text{G}_{\text{gs}}(\phi_\mathrm{t}, \theta_\mathrm{t})$ denotes the gain of the GS antenna at the departure angles $(\phi_\mathrm{t}, \theta_\mathrm{t})$, and $d_{\mathrm{3D}}$ is the 3D Euclidean distance.
 
\section{Simulated Radiation Pattern of UAV-Antenna}
\label{sec:app2}
\begin{figure}[t!]
    \centering
    \begin{subfigure}{0.48\textwidth}
        \centering
        \includegraphics[width=\linewidth]{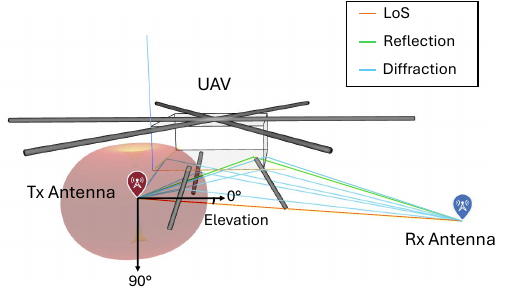}
        \caption{Standalone receiver antenna}
    \end{subfigure}
    \hfill
    \begin{subfigure}{0.48\textwidth}
        \centering
        \includegraphics[width=\linewidth]{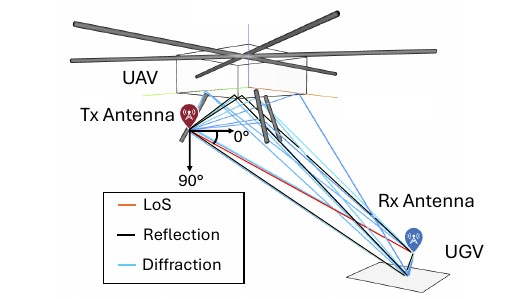}
        \caption{Receiver antenna mounted on UGV}
    \end{subfigure}
    \caption{MATLAB ray tracing simulation environment for measuring the effective radiation pattern of a UAV antenna: (a) with a standalone receiver antenna, (b) with a receiver antenna mounted on a UGV.}
    \label{fig:mat_simu_env}
    \vspace{-4mm}
\end{figure}

To validate our hypotheses regarding \raisebox{.5pt}{\textcircled{\raisebox{-.9pt}{4}}} and \raisebox{.5pt}{\textcircled{\raisebox{-.9pt}{3}}} observed in Fig.~\ref{fig:ant_patt_elev} (i.e., at medium and lower elevation angles, respectively), we conduct a ray tracing simulation using MATLAB. In a 3D environment, we model a UAV with six propeller arms, three landing gears, and a metal cuboid measuring \(50\)~cm by \(50\)~cm by \(20\)~cm, as shown in Fig.~\ref{fig:mat_simu_env}(a).
The UAV antenna, oriented downward, is positioned \(20\)~cm below the UAV's center and \(40\)~cm to the left, providing a lateral offset. This positioning is suitable for reflecting rays in the opposite horizontal direction, utilizing the bottom surface of the cuboid. 
The design is inspired by the UAV-based spectrum sensor, where the cuboid represents the portable node attached beneath the UAV. To measure the radiation pattern of the UAV-mounted antenna, an isotropic receiver antenna is positioned at approximately \(500\)~m from the UAV, with elevation angles ranging from \(0^\circ\) to \(90^\circ\) relative to the UAV antenna. Additionally, we repeated the simulation with an extra metallic surface measuring \(100\)~cm by \(100\)~cm, placed \(12\)~cm beneath the receiver antenna, as shown in Fig.~\ref{fig:mat_simu_env}(b). The azimuth angle range spans \(20^\circ\), centered on the direction where reflections from the UAV and UGV surfaces are most prominent. For ray tracing, the maximum number of reflections and diffractions was set to 2 and 1, respectively.

\begin{figure}[t!]
    \centering
    \begin{subfigure}{0.40\textwidth}
        \centering
        \includegraphics[width=\linewidth]{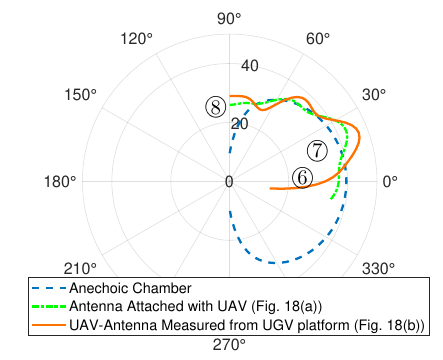}
    \end{subfigure}
    \vspace{-2mm}
    \caption{The elevation radiation pattern of a UAV-mounted antenna, which initially exhibits a dipole pattern when measured standalone in an anechoic chamber. The effective radiation pattern of the UAV-mounted antenna is computed using \eqref{eqn:A_uav_calc} and \eqref{eq:effective_patt_calc} within a MATLAB simulation environment, as shown in Fig.~\ref{fig:mat_simu_env}(a) and Fig.~\ref{fig:mat_simu_env}(b).}
    \label{fig:elev_patt_matlab_simu}
\end{figure}

The pattern of received power strength at various elevation angles is illustrated in Fig.~\ref{fig:elev_patt_matlab_simu}.
At lower elevation angles, highlighted by \raisebox{.5pt}{\textcircled{\raisebox{-.9pt}{6}}}, the measurement environments of both Fig.~\ref{fig:mat_simu_env}(a) and Fig.~\ref{fig:mat_simu_env}(b) show reduced antenna gain. Further analysis reveals that the lower gain in Fig.~\ref{fig:mat_simu_env}(a) is primarily due to destructive interference of the reflected ray, while in Fig.~\ref{fig:mat_simu_env}(b), the gain is further diminished by LoS blocking.
At medium elevation angles, highlighted by \raisebox{.5pt}{\textcircled{\raisebox{-.9pt}{7}}}, the effective antenna gain increases due to constructive interference of the reflected rays. In Fig.~\ref{fig:mat_simu_env}(b), the number of reflected rays increases, leading to an even greater gain. Because of the limited size of the UAV cuboid, the reflected rays from the UAV disappear beyond \(30^\circ\) elevation, and the effective pattern in Fig.~\ref{fig:mat_simu_env}(a) begins to resemble the anechoic chamber patterns after \(30^\circ\). In contrast, the size and placement of the additional UGV surface in Fig.~\ref{fig:mat_simu_env}(b) continue to generate reflected rays at higher elevation angles, leading to periodic constructive and destructive interference as the elevation angle increases.
Finally, at around the \(90^\circ\) elevation angle, the effective antenna gain is significantly higher than that of the anechoic chamber. This is because the diffracted rays reach the \(90^\circ\) elevation angle, becoming stronger than the LoS ray. For a dipole pattern, the antenna gain at \(90^\circ\) elevation angle is extremely low, so the diffracted rays dominate.

\section{Tri-phasic REM Accuracy across Altitudes: Two Additional Case Studies}
\label{sec:app3}
\subsection{Multi-band dataset}
\begin{figure}[!t]
    \centering
        \includegraphics[width=\linewidth]{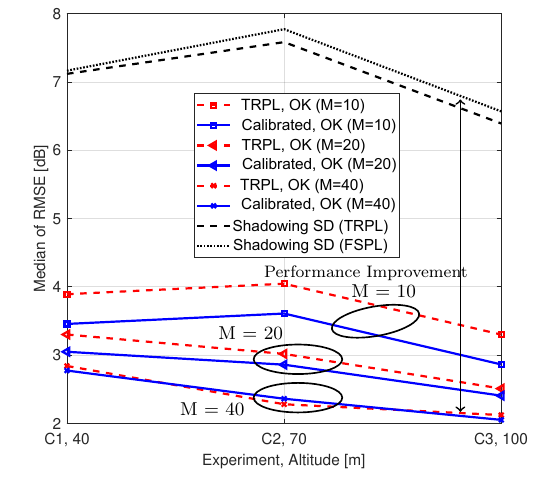}
    \caption{Performance of RSRP prediction across three experiments with increasing altitudes from the multi-band dataset. The results are presented for TRPL with OK, both with and without APC (solid blue and dashed red lines, respectively).}
    \label{fig:altitudes_cole_3}
    \vspace{-5mm}
\end{figure}

Fig.~\ref{fig:altitudes_cole_3} presents the results of the median of RMSE for three experiments from the AERPAW multi-band dataset, with altitudes \(40 \, \text{m}\), \(70 \, \text{m}\), and \(100 \, \text{m}\).
As observed, the estimated SF standard deviation $\sigma$ is highest at the intermediate altitude of 70~m. 
Given that this dataset only includes three altitude samples, it serves as a partial observation of the previously identified tri-phasic behavior of $\sigma$. If additional measurements were available at altitudes closer to the ground, we would expect a higher $\sigma$ compared to the 40~m case, driven by the increased presence of ground-level multipath reflections and clutter. Thus, the current three-point profile likely captures the mid-altitude peak and the subsequent transition toward the high-altitude LoS-dominant regime. The reconstruction error curves using OK also follow the trend of \(\sigma\).

\subsection{AFAR dataset}
\begin{figure}[t!]
        \centering
        \vspace{3mm}
        \includegraphics[width=\linewidth]{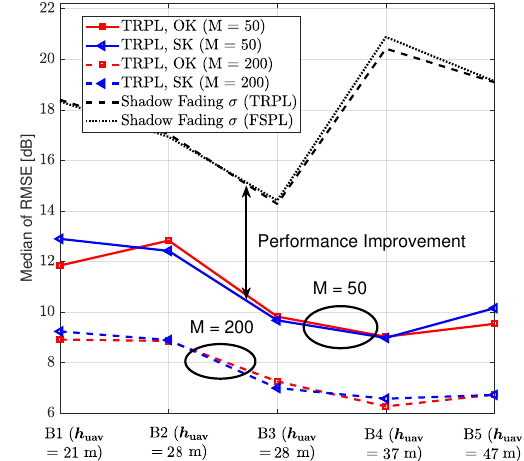}
    \caption{Performance of RSRP prediction across five experiments with non-decreasing altitudes from the AFAR dataset. The results are presented for TRPL with OK and SK.}
    \label{fig:trajectories_all_5}
    \vspace{-4mm}
\end{figure}

Fig.~\ref{fig:trajectories_all_5} presents the median RMSE results using TRPL with OK and SK for the experimental AFAR dataset.
The experiments are arranged in ascending order of UAV altitudes, ranging from 21~m to 47~m. Notably, in this dataset, the GS was a UGV with an antenna height of approximately 1.5~m. The estimated SF standard deviation $\sigma$ across these experiments exhibits a tri-phasic behavior: $\sigma$ initially decreases from 21~m to 28~m, increases at 37~m, and subsequently declines at 47~m. While these experiments utilized distinct 2D trajectories (as illustrated in Fig.~\ref{fig:trajectories_afar}), the significant fluctuations in $\sigma$ are predominantly attributed to the variations in UAV altitude rather than the specific horizontal paths. Interestingly, the reconstructed REM error trends do not strictly mirror the $\sigma$ profile, particularly at the 37~m and 47~m altitudes. This divergence occurs because higher altitudes result in stronger spatial correlation; consequently, the performance gains provided by OK and SK are more pronounced at these heights, decoupling the reconstruction error trend from the underlying SF standard deviation \(\sigma\).

\section{Inter-trajectory Antenna Calibration}
\label{sec:app4}
\begin{figure}[t!]
\centerline{\includegraphics[width=\linewidth,trim={0cm 0cm 0cm 0.25cm},clip]{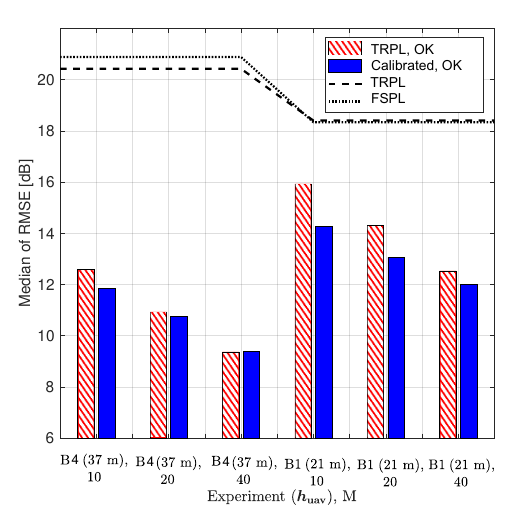}}
\caption{Comparison of the performance of the TRPL and calibrated TRPL across two experiments from the AFAR dataset. The calibrated model shows improved performance at lower sampling rate, \(\mathrm{M}\).}
\label{fig:trajectories_2_calibrated}
\vspace{-4mm}
\end{figure}

Fig.~\ref{fig:trajectories_2_calibrated} presents the REM reconstruction performance of the APC-enhanced TRPL compared to the baseline TRPL using the AFAR dataset.
The AFAR dataset comprises five experiments of varied trajectories as shown in Fig.~\ref{fig:trajectories_afar}. Calibrating the antenna pattern using data from a different trajectory seems more challenging than calibrating from a different altitude of the same trajectory. This is because changing the altitude of the same trajectory only affects the elevation angle of signal reception, while varying the trajectory also alters the azimuth angle. Therefore, we carefully select experiments for APC. Specifically, experiment B5 is used for calibration when testing on experiment B4, due to the similarity between their trajectories. Similarly, we calibrate the effective antenna pattern using data from experiments B2 and B3 to test on experiment B1. The results of REM reconstruction with the calibrated pattern for experiments B4 and B1 are shown in Fig.~\ref{fig:trajectories_2_calibrated}. The results show that the effective radiation pattern is more helpful when the number of measurements \(\mathrm{M}\) is smaller, similar to previous cases. The maximum improvement, approximately 2 dB, is observed for experiment B1 with \(\mathrm{M} = 10 \).

\begin{figure}[t!]
    \centering
    \begin{subfigure}{0.48\textwidth}
        \centering
        \includegraphics[width=\linewidth]{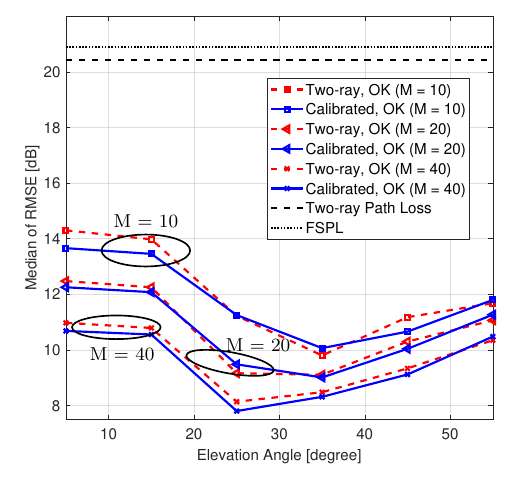}
        \caption{Experiment B4}
    \end{subfigure}
    \hfill
    \begin{subfigure}{0.48\textwidth}
        \centering
        \includegraphics[width=\linewidth]{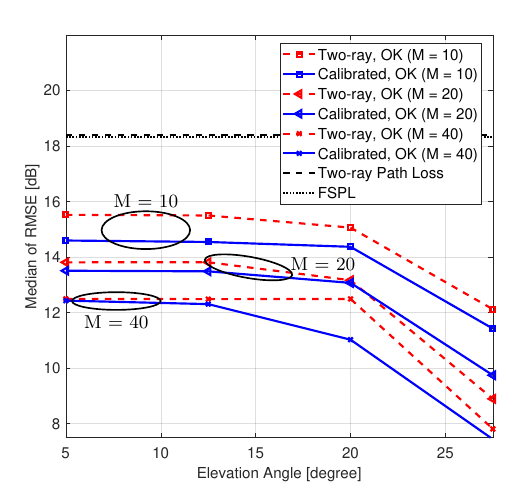}
        \caption{Experiment B1}
    \end{subfigure}
    \caption{REM reconstruction error across elevation angles for TRPL and APC-enhanced TRPL, evaluated on (a) experiment B4 and (b) experiment B1. APC yields improvement for elevation angles up to 20\(^\circ\). The results are obtained using OK at \( \mathrm{R} = 70 \, \text{m} \).}
    \label{fig:trajectories_2_reason}
    \vspace{-4mm}
\end{figure}

To further investigate the cases, we plot the median RMSE across elevation angles for experiments B4 and B1 in Fig.~\ref{fig:trajectories_2_reason}.
The comparative plots for the TRPL and APC-enhanced TRPL suggest that the error is reduced for lower elevation angles (less than \(20^\circ\)) in both cases, which corresponds to learning the deep fades at these elevation angles, as demonstrated in Fig.~\ref{fig:ant_patt_elev}(c) and Fig.~\ref{fig:ant_patt_elev}(d). Experiment B1 in Fig.~\ref{fig:trajectories_2_reason}(b) does not have samples above \(30^\circ\). However, the results from experiment B4 in Fig.~\ref{fig:trajectories_2_reason}(a) show that APC performance degrades at larger elevation angles relative to the baseline. This suggests that the radiation pattern has not been properly adjusted for higher elevation angles, likely due to insufficient sampling in training experiment B5 at the azimuth and elevation angles covered by experiment B4. In such cases, the system may revert to the baseline anechoic chamber radiation pattern rather than applying APC.

\bibliographystyle{IEEEtran}
\bibliography{ref}

@STRING{IEEE_comm = "IEEE Trans. Commun."}

@STRING{IEEE_cog_comm = "IEEE Trans. Cognitive Commun. Netw."}

@STRING{IEEE_open_comm = "IEEE Open J. Commun. Societ."}

@STRING{IEEE_veh = "IEEE Trans. Vehi. Technol."}

@STRING{IEEE_veh_mag = "IEEE Vehi. Technol. Mag."}

@STRING{IEEE_sensors = "IEEE Sensors J."}

@STRING{IEEE_wireless = "IEEE Trans. Wireless Commun."}

@STRING{IEEE_ICC = "Proc. IEEE Int. Conf. Commun."}

@STRING{IEEE_ICC_w = "Proc. IEEE Int. Conf. Commun. Workshops (ICC Workshops)"}

@STRING{IEEE_things = "IEEE Internet of Things J."}

@STRING{IEEE_WPMC = "Proc. Int. Symp. Wireless Personal Multimedia Commun. (WPMC)"}

@STRING{IEEE_PIMRC = "Proc. IEEE Annual Int. Symp. Personal, Indoor, Mobile Radio Commun. (PIMRC)"}

@STRING{IEEE_sig_process = "IEEE Trans. Sig. Process."}

@STRING{IEEE_wisnet = "Proc. IEEE Topical Conf. Wireless Sensors Sensor Netw. (WiSNet)"}

@STRING{IEEE_dyspan = "Proc. IEEE Int. Symp. Dynamic Spectrum Access Netw. (DySPAN)"}

@article{sun2022propagation,
  title={Propagation map reconstruction via interpolation assisted matrix completion},
  author={Sun, Hao and Chen, Junting},
  journal=IEEE_sig_process,
  volume={70},
  pages={6154--6169},
  year={2022}
}

@article{maeng2023kriging,
  title={{Kriging-Based 3-D Spectrum Awareness for Radio Dynamic Zones Using Aerial Spectrum Sensors}},
  author={Maeng, Sung Joon and Ozdemir, Ozgur and G{\"u}ven{\c{c}}, {\.I}smail and Sichitiu, Mihail L},
  journal=IEEE_sensors,
  volume={24},
  number={6},
  pages={9044--9058},
  year={2024}
}

@article{sato2017kriging,
  title={Kriging-based interference power constraint: Integrated design of the radio environment map and transmission power},
  author={Sato, Koya and Fujii, Takeo},
  journal=IEEE_cog_comm,
  volume={3},
  number={1},
  pages={13--25},
  year={2017}
}

@article{szyszkowicz2010feasibility,
  title={On the feasibility of wireless shadowing correlation models},
  author={Szyszkowicz, Sebastian S and Yanikomeroglu, Halim and Thompson, John S},
  journal=IEEE_veh,
  volume={59},
  number={9},
  pages={4222--4236},
  year={2010}
}

@article{zhang2020spectrum,
  title={Spectrum cartography via coupled block-term tensor decomposition},
  author={Zhang, Guoyong and Fu, Xiao and Wang, Jun and Zhao, Xi-Le and Hong, Mingyi},
  journal=IEEE_sig_process,
  volume={68},
  pages={3660--3675},
  year={2020},
}

@misc{sa_1400,
author = {{Octane Wireless}},
title = {{SA-1400-5900 Data Sheet}},
url = {https://www.octanewireless.com/product/sa-1400-5900-tri-band-stub-antenna/},
}

@article{mobilemarkrmwb1,
    author = {Mobile Mark, Inc},
    title = {{RM-WB1} Series Radiation Pattern},
    url = {[Online]. Available: https://www.mobilemark.com/product/rm-wb1/}
}

@book{cressie2015statistics,
  title={{Statistics for Spatial Data}},
  author={Cressie, Noel},
  year={2015},
  publisher={John Wiley \& Sons}
}

@ARTICLE{9061157,
  author={Marojevic, Vuk and Guvenc, Ismail and Dutta, Rudra and Sichitiu, Mihail L. and Floyd, Brian A.},
  journal=IEEE_veh_mag, 
  title={Advanced Wireless for Unmanned Aerial Systems: {5G} Standardization, Research Challenges, and {AERPAW} Architecture}, 
  year={2020},
  volume={15},
  number={2},
  pages={22-30},
}

@data{IEEEDataPort,
doi = {10.21227/mma2-0t93},
url = {https://dx.doi.org/10.21227/mma2-0t93},
author = {Maeng, Sung Joon and Ozdemir, Ozgur and Guvenc, Ismail and Sichitiu, Mihail},
howpublished = {IEEE Dataport},
title = {{3D antenna pattern measurement for AERPAW experiments}},
year = {2023} }

@data{IEEEDataPort_2,
doi = {10.21227/0p43-0d72},
url = {https://dx.doi.org/10.21227/0p43-0d72},
author = {Maeng, Sung Joon and Ozdemir, Ozgur and Guvenc, Ismail and Sichitiu, Mihail and Dutta, Rudra},
howpublished = {IEEE Dataport},
title = {{LTE I/Q} Measurement by {AERPAW} Platform for Air-to-Ground Propagation Modeling},
year = {2022} }

@data{IEEEDataPort_3,
doi = {10.5061/dryad.18931zd4g},
url = {https://doi.org/10.5061/dryad.18931zd4g},
author = {Gurses, Anil and Fahim Raouf, Amir Hossein and Maeng, Sung Joon and Ozdemir, Ozgur and Guvenc, Ismail and Sichitiu, Mihail},
howpublished = {IEEE Dataport},
title = {{AERPAW Find-a-Rover (AFAR) Challenge in December 2023}},
year = {2024} }

@data{IEEEDataPort_4,
doi = {10.5061/dryad.2z34tmpvv},
url = {https://doi.org/10.5061/dryad.2z34tmpvv},
author = {Dickerson, Cole and Fahim Raouf, Amir Hossein and Ozdemir, Ozgur and Guvenc, Ismail and Sichitiu, Mihail},
howpublished = {IEEE Dataport},
title = {{AERPAW UAV-based signal data collected at varying altitudes and sampling rates for wireless communication studies}},
year = {2024} }

@article{wackernagel2003ordinary,
  title={Ordinary Kriging},
  author={Wackernagel, Hans and Wackernagel, Hans},
  journal={{Multivariate Geostatistics: An Introduction with Applications}},
  pages={79--88},
  year={2003},
  publisher={Springer}
}

@article{olea1999simple,
  title={Simple Kriging},
  author={Olea, Ricardo A and Olea, Ricardo A},
  journal={{Geostatistics for Engineers and Earth Scientists}},
  pages={7--30},
  year={1999},
  publisher={Springer}
}

@article{olea1994fundamentals,
  title={Fundamentals of semivariogram estimation, modeling, and usage},
  author={Olea, Ricardo A},
  year={1994},
  publisher={AAPG Special Volumes}
}

@book{casella2021statistical,
  title={{Statistical Inference}},
  author={Casella, George and Berger, Roger L},
  year={2021},
  publisher={Cengage Learning}
}

@article{Shabat2023,
year={2024},
author={Shabat, Gil},
  publisher={MATLAB Central File Exchange},
title={Matrix completion using nuclear norm, spectral norm or weighted nuclear norm minimization},
  url = {https://www.mathworks.com/matlabcentral/fileexchange/50056-matrix-completion-using-nuclear-norm-spectral-norm-or-weighted-nuclear-norm-minimization}
}

@article{badi2020experimentally,
  title={Experimentally analyzing diverse antenna placements and orientations for UAV communications},
  author={Badi, Mahmoud and Wensowitch, John and Rajan, Dinesh and Camp, Joseph},
  journal=IEEE_veh,
  volume={69},
  number={12},
  pages={14989--15004},
  year={2020}
}

@inproceedings{badi2019experimental,
  title={Experimental evaluation of antenna polarization and elevation effects on drone communications},
  author={Badi, Mahmoud and Wensowitch, John and Rajan, Dinesh and Camp, Joseph},
  booktitle={Proc. Int. ACM Conf. Modeling, Analysis Simulation Wireless Mobile Syst.},
  pages={211--220},
  year={2019},
  month={November},
  address={Miami, USA}
}

@article{sun2017air,
  title={{Air-ground channel characterization for unmanned aircraft systems—Part IV: Airframe shadowing}},
  author={Sun, Ruoyu and Matolak, David W and Rayess, William},
  journal=IEEE_veh,
  volume={66},
  number={9},
  pages={7643--7652},
  year={2017}
}

@article{wang2023sparse,
  title={Sparse {Bayesian} learning-based {3-D} radio environment map construction—Sampling optimization, scenario-dependent dictionary construction, and sparse recovery},
  author={Wang, Jie and Zhu, Qiuming and Lin, Zhipeng and Wu, Qihui and Huang, Yang and Cai, Xuezhao and Zhong, Weizhi and Zhao, Yi},
  journal=IEEE_cog_comm,
  volume={10},
  number={1},
  pages={80--93},
  year={2023},
  publisher={IEEE}
}

@article{zhou2023accurate,
  title={Accurate spectrum map construction for spectrum management through intelligent frequency-spatial reasoning},
  author={Zhou, Fuhui and Wang, Chenyue and Wu, Guangyu and Wu, Yuhang and Wu, Qihui and Al-Dhahir, Naofal},
  journal=IEEE_comm,
  volume={71},
  number={7},
  pages={3932--3945},
  year={2023},
  publisher={IEEE}
}

@article{shen2019uav,
  title={{UAV-based 3D spectrum sensing in spectrum-heterogeneous networks}},
  author={Shen, Feng and Ding, Guoru and Wang, Zheng and Wu, Qihui},
  journal=IEEE_veh,
  volume={68},
  number={6},
  pages={5711--5722},
  year={2019},
  publisher={IEEE}
}

@article{wei2022three,
  title={{Three-dimensional spectrum occupancy measurement using UAV: Performance analysis and algorithm design}},
  author={Wei, Zhiqing and Yao, Rubing and Kang, Jie and Chen, Xu and Wu, Huici},
  journal=IEEE_sensors,
  volume={22},
  number={9},
  pages={9146--9157},
  year={2022}
}

@inproceedings{romero2020aerial,
  title={{Aerial spectrum surveying: Radio map estimation with autonomous UAVs}},
  author={Romero, Daniel and Shrestha, Raju and Teganya, Yves and Chepuri, Sundeep Prabhakar},
  booktitle={Proc. IEEE Int. Workshop Machine Learning Signal Process. (MLSP)},
  pages={1--6},
  year={2020},
  month={September}
}

@article{du2021uav,
  title={{UAV-assisted three-dimensional spectrum mapping driven by spectrum data and channel model}},
  author={Du, Xiaofu and Zhu, Qiuming and Ding, Guoru and Li, Jie and Wu, Qihui and Lan, Tianxu and Lin, Zhipeng and Zhong, Weizhi and Han, Lu},
  journal={Symmetry},
  volume={13},
  number={12},
  pages={2308},
  year={2021},
  publisher={MDPI}
}

@article{shrestha2022spectrum,
  title={Spectrum surveying: Active radio map estimation with autonomous UAVs},
  author={Shrestha, Raju and Romero, Daniel and Chepuri, Sundeep Prabhakar},
  journal=IEEE_wireless,
  volume={22},
  number={1},
  pages={627--641},
  year={2022},
  publisher={IEEE}
}

@article{ivanov2022uav,
  title={{UAV-based volumetric measurements toward radio environment map construction and analysis}},
  author={Ivanov, Antoni and Muhammad, Bilal and Tonchev, Krasimir and Mihovska, Albena and Poulkov, Vladimir},
  journal={Sensors},
  volume={22},
  number={24},
  pages={9705},
  year={2022},
  publisher={MDPI}
}

@inproceedings{zhu2022demo,
  title={{Demo abstract: An UAV-based 3D spectrum real-time mapping system}},
  author={Zhu, Qiuming and Zhao, Yi and Huang, Yang and Lin, Zhipeng and Wang, Lu HanJie and Bai, Yunpeng and Lan, Tianxu and Zhou, Fuhui and Wu, Qihui},
  booktitle={Proc. IEEE Conf. Computer Commun. Workshops (INFOCOM WKSHPS)},
  pages={1--2},
  year={2022},
  month={May}
}

@inproceedings{zeng2022uav,
  title={{UAV-aided joint radio map and 3D environment reconstruction using deep learning approaches}},
  author={Zeng, Pengxi and Chen, Junting},
  booktitle=IEEE_ICC,
  pages={5341--5346},
  year={2022},
  month={May},
  address={Seoul, South Korea}
}

@article{liu2023uav,
  title={{UAV-aided radio map construction exploiting environment semantics}},
  author={Liu, Wenjie and Chen, Junting},
  journal=IEEE_wireless,
  volume={22},
  number={9},
  pages={6341--6355},
  year={2023}
}

@inproceedings{tsukamoto2018highly,
  title={Highly accurate radio environment mapping method based on transmitter localization and spatial interpolation in urban {LoS/NLoS} scenario},
  author={Tsukamoto, Kenta and Kitsunezuka, Masaki and Kunihiro, Kazuaki},
  booktitle=IEEE_WiSNet,
  pages={5--7},
  year={2018},
  month={January},
  address={California, USA}
}

@article{mao2022machine,
  title={{Machine-learning-based 3-D channel modeling for U2V mmWave communications}},
  author={Mao, Kai and Zhu, Qiuming and Song, Maozhong and Li, Hanpeng and Ning, Benzhe and Pedersen, Gert Fr{\o}lund and Fan, Wei},
  journal=IEEE_things,
  volume={9},
  number={18},
  pages={17592--17607},
  year={2022}
}

@article{pesko2015indirect,
  title={The indirect self-tuning method for constructing radio environment map using omnidirectional or directional transmitter antenna},
  author={Pesko, Marko and Javornik, Toma{\v{z}} and Vidmar, Luka and Ko{\v{s}}ir, Andrej and {\v{S}}tular, Mitja and Mohor{\v{c}}i{\v{c}}, Mihael},
  journal={EURASIP J. Wireless Commun. Netw.},
  volume={2015},
  pages={1--12},
  year={2015},
  publisher={Springer}
}

@article{chaves2023deeprem,
  title={{DeepREM: Deep-learning-based radio environment map estimation from sparse measurements}},
  author={Chaves-Villota, Andrea and Viteri-Mera, Carlos A},
  journal={IEEE Access},
  volume={11},
  pages={48697--48714},
  year={2023}
}

@article{xia2020radio,
  title={{Radio environment map construction by adaptive ordinary Kriging algorithm based on affinity propagation clustering}},
  author={Xia, Haiyang and Zha, Song and Huang, Jijun and Liu, Jibin},
  journal={Int. J. Distributed Sensor netw.},
  volume={16},
  number={5},
  pages={1550147720922484},
  year={2020},
  publisher={SAGE Publications Sage UK: London, England}
}

@inproceedings{gajewski2018propagation,
  title={Propagation models in radio environment map design},
  author={Gajewski, Piotr},
  booktitle={Proc. IEEE Baltic URSI Symposium (URSI)},
  pages={234--237},
  address={Poznan, Poland},
  year={2018},
  month={May}
}

@inproceedings{farnham2016radio,
  title={Radio environment map techniques and performance in the presence of errors},
  author={Farnham, Tim},
  booktitle=IEEE_PIMRC,
  pages={1--6},
  year={2016},
  address={Valencia, Spain},
  month={September}
}

@article{ivanov2023limited,
  title={Limited sampling spatial interpolation evaluation for {3D} radio environment mapping},
  author={Ivanov, Antoni and Tonchev, Krasimir and Poulkov, Vladimir and Manolova, Agata and Vlahov, Atanas},
  journal={Sensors},
  volume={23},
  number={22},
  pages={9110},
  year={2023},
  publisher={MDPI}
}

@inproceedings{ivanov2024calibration,
  title={{Calibration of Statistical Interpolation through Anechoic Chamber Measurements for 3D REM Reconstruction}},
  author={Ivanov, Antoni and Vlahov, Atanas and Poulkov, Vladimir and Yioultsis, Traianos and Zaharis, Zaharias},
  booktitle=IEEE_WPMC,
  pages={1--5},
  year={2024},
  month={November},
  address={India}
}

@article{ivanov2024deep,
  title={Deep Learning for Reduced Sampling Spatial {3D REM} Reconstruction},
  author={Ivanov, Antoni and Tonchev, Krasimir and Poulkov, Vladimir and Manolova, Agata},
  journal=IEEE_open_comm,
  year={2024},
  publisher={IEEE}
}

@inproceedings{rahman20243d,
  title={{3D Spectrum Awareness for Radio Dynamic Zones Using Kriging and Matrix Completion}},
  author={Rahman, Mushfiqur and Maeng, Sung Joon and G{\"u}ven{\c{c}}, {\.I}smail and Wong, Chau-Wai},
  booktitle=IEEE_dyspan,
  pages={439--446},
  year={2024},
  month={May},
  address={Washington, DC, USA}
}

@article{wang2024sparse,
  title={{Sparse Bayesian learning-based hierarchical construction for 3D radio environment maps incorporating channel shadowing}},
  author={Wang, Jie and Zhu, Qiuming and Lin, Zhipeng and Chen, Junting and Ding, Guoru and Wu, Qihui and Gu, Guochen and Gao, Qianhao},
  journal=IEEE_wireless,
  volume={23},
  number={10},
  pages={14560--14574},
  year={2024},
  publisher={IEEE}
}

@article{shen20213d,
  title={{3D} compressed spectrum mapping with sampling locations optimization in spectrum-heterogeneous environment},
  author={Shen, Feng and Wang, Zheng and Ding, Guoru and Li, Kezhi and Wu, Qihui},
  journal=IEEE_wireless,
  volume={21},
  number={1},
  pages={326--338},
  year={2021},
  publisher={IEEE}
}

@article{alobaidy2025empirical,
  title={Empirical {3-D} Channel Modeling for Cellular-Connected {UAVs}: A Triple-Layer Machine Learning Approach},
  author={Alobaidy, Haider AH and Behjati, Mehran and Nordin, Rosdiadee and Zulkifley, Muhammad Aidiel and Abdullah, Nor Fadzilah and Salleh, Nur Fahimah Mat},
  journal=IEEE_open_comm,
  volume={6},
  pages={9908--9925},
  year={2025},
  publisher={IEEE}
}

@article{wang2025robust,
  title={A robust learning framework for spatial-temporal-spectral radio map prediction},
  author={Wang, Lei and Hu, Jun and Jiang, Dan and Chen, Zengping},
  journal={Expert Syst. Appl.},
  pages={129351},
  year={2025},
  publisher={Elsevier}
}

@inproceedings{semkin2021lightweight,
  title={{Lightweight UAV-based measurement system for air-to-ground channels at 28 GHz}},
  author={Semkin, Vasilii and Kang, Seongjoon and Haarla, Jaakko and Xia, William and Huhtinen, Ismo and Geraci, Giovanni and Lozano, Angel and Loianno, Giuseppe and Mezzavilla, Marco and Rangan, Sundeep},
  booktitle=IEEE_PIMRC,
  pages={848--853},
  year={2021},
  month={September}
}

@article{hu20233d,
  title={{3D} radio map reconstruction based on generative adversarial networks under constrained aircraft trajectories},
  author={Hu, Tianyu and Huang, Yang and Chen, Junting and Wu, Qihui and Gong, Zhiren},
  journal=IEEE_veh,
  volume={72},
  number={6},
  pages={8250--8255},
  year={2023}
}

@inproceedings{krijestorac2021spatial,
  title={Spatial signal strength prediction using {3D} maps and deep learning},
  author={Krijestorac, Enes and Hanna, Samer and Cabric, Danijela},
  booktitle=IEEE_ICC,
  pages={1--6},
  year={2021},
  month={June}
}

@article{levie2021radiounet,
  title={{RadioUNet: Fast radio map estimation with convolutional neural networks}},
  author={Levie, Ron and Yapar, {\c{C}}a{\u{g}}kan and Kutyniok, Gitta and Caire, Giuseppe},
  journal=IEEE_wireless,
  volume={20},
  number={6},
  pages={4001--4015},
  year={2021},
  publisher={IEEE}
}

@inproceedings{locke2023radio,
  title={Radio map estimation with deep dual path autoencoders and skip connection learning},
  author={Locke, William and Lokhmachev, Nikita and Huang, Yan and Li, Xinrong},
  booktitle=IEEE_PIMRC,
  pages={1--6},
  year={2023},
  address={Toronto, Canada},
  month={September}
}

@article{teganya2021deep,
  title={Deep completion autoencoders for radio map estimation},
  author={Teganya, Yves and Romero, Daniel},
  journal=IEEE_wireless,
  volume={21},
  number={3},
  pages={1710--1724},
  year={2021},
  publisher={IEEE}
}

@article{wang2024radiodiff,
  title={{RadioDiff: An effective generative diffusion model for sampling-free dynamic radio map construction}},
  author={Wang, Xiucheng and Tao, Keda and Cheng, Nan and Yin, Zhisheng and Li, Zan and Zhang, Yuan and Shen, Xuemin},
  journal=IEEE_cog_comm,
  volume={11},
  number={2},
  pages={738--750},
  year={2024},
  publisher={IEEE}
}

@article{zhang2023rme,
  title={{RME-GAN: A learning framework for radio map estimation based on conditional generative adversarial network}},
  author={Zhang, Songyang and Wijesinghe, Achintha and Ding, Zhi},
  journal=IEEE_things,
  volume={10},
  number={20},
  pages={18016--18027},
  year={2023},
  publisher={IEEE}
}

@article{zhang2024generative,
  title={{Generative AI on SpectrumNet: An open benchmark of multiband 3-D radio maps}},
  author={Zhang, Shuhang and Jiang, Shuai and Lin, Wanjie and Fang, Zheng and Liu, Kangjun and Zhang, Hongliang and Chen, Ke},
  journal=IEEE_cog_comm,
  volume={11},
  number={2},
  pages={886--901},
  year={2024},
  publisher={IEEE}
}

@article{masrur2025collection,
  title={Collection: {UAV}-Based {RSS} Measurements from the {AFAR} Challenge in Digital Twin and Real-World Environments},
  author={Masrur, Saad and Ozdemir, Ozgur and Gurses, Anil and Guvenc, Ismail and Sichitiu, Mihail L. and Dutta, Rudra and Mushi, Magreth and Zajkowski, homas and Dickerson, Cole and Reddy, Gautham and Villar, Sergio Vargas and Wong, Chau-Wai and Chatterjee, Baisakhi and Chaudhari, Sonali and Li, Zhizhen and Liu, Yuchen and Kudyba, Paul and Sun, Haijian and Mandapaka, Jaya Sravani and Namuduri, Kamesh and Wang, Weijie and Fund, Fraida},
  journal={IEEE Data Descrip.},
  year={2025}
}

@article{gurses2025digital,
  title={Digital twins and testbeds for supporting {AI} research with autonomous vehicle networks},
  author={Gurses, Anil and Reddy, Gautham and Masrur, Saad and Ozdemir, Ozgur and Guvenc, Ismail and Sichitiu, Mihail L and Sahin, Alphan and Alkhateeb, Ahmed and Mushi, Magreth and Dutta, Rudra},
  journal={IEEE Commun. Mag.},
  volume={63},
  number={4},
  pages={56--62},
  year={2025}
}

@inproceedings{masrur2025bridging,
  title={Bridging simulation and reality: A {3D} clustering-based deep learning model for {UAV}-based {RF} source localization},
  author={Masrur, Saad and G{\"u}ven{\c{c}}, Ismail},
  booktitle=IEEE_ICC_w,
  pages={953--958},
  year={2025},
  month={June},
  address={Montreal, Canada}
}

@inproceedings{reddy2026transforem,
  title={{TransfoREM: Transformer} aided {3D} Radio Environment Mapping},
  author={Reddy, Gautham and Guvenc, Ismail and Sichitiu, Mihail L and Bhuyan, Arupjyoti and Petersen, Bryton and Abrahamson, Jason},
  booktitle=IEEE_ICC,
  year={2026},
  month={May},
  address={Glasgow, Scotland, UK},
  note={to appear}
}

\end{document}